 \documentclass[final,3p,times,twocolumn]{elsarticle}

\usepackage{amssymb}

\biboptions{sort&compress}

\journal{NIM B346(2015)26}

\begin{document}

\begin{frontmatter}



\title{Activation cross-sections of proton induced reactions on $^{nat}$Sm up to 65 MeV}


\author[1]{F. T\'ark\'anyi}
\author[2]{A. Hermanne}
\author[1]{S. Tak\'acs}
\author[1]{F. Ditr\'oi\corref{*}}
\author[3]{A.V. Ignatyuk}
\cortext[*]{Corresponding author: ditroi@atomki.hu}

\address[1]{Institute for Nuclear Research, Hungarian Academy of Sciences (ATOMKI),  Debrecen, Hungary}
\address[2]{Cyclotron Laboratory, Vrije Universiteit Brussel (VUB), Brussels, Belgium}
\address[3]{Institute of Physics and Power Engineering (IPPE), Obninsk, Russia}

\begin{abstract}
Activation cross  sections for proton induced reactions on Sm  are presented for the first time for $^{nat}$Sm(p,xn)$^{154,152m2,152m1,152g,150m,150g,149,148,147,146,145}$Eu, $^{nat}$Sm(p,x)$^{153,145}$Sm, $^{nat}$Sm(p,x)$^{151,150,149,148g,148m,146,144,143}$Pm and $^{nat}$Sm(p,x)$^{141}$Nd up to 65 MeV. The cross sections were measured via activation method by using a stacked-foil irradiation technique and high resolution gamma ray spectroscopy. The results were compared with results of the nuclear reaction codes ALICE, EMPIRE and TALYS (results taken from TENDL libraries). Integral yields of the activation products were calculated from the excitation functions.
\end{abstract}

\begin{keyword}
proton irradiation\sep samarium target\sep europium, samarium, promethium and neodymium radio-isotopes\sep cross-section measurement\sep yield calculation

\end{keyword}

\end{frontmatter}


\section{Introduction}
\label{1}
At present, targeted radiotherapy (TR) is acknowledged to have great potential in oncology. A large list of interesting radionuclides was identified, including several radioisotopes of lanthanides, amongst them $^{145}$Sm and $^{153}$Sm. $^{145}$Sm plays an important role in brachytherapy and $^{153}$Sm is used in bone pain palliation treatments. The nuclear data for their production around a reactor were compiled and evaluated in two IAEA nuclear data libraries \cite{1, 2}. We investigated their possible production at a cyclotron using a 50 MeV deuteron beam on samarium target \cite{3, 4} in an effort to fulfill the requirements of nuclear medicine in medical radioisotope production without relying on nuclear reactors \cite{5, 6}. A short overview of applications and comparison of possible production routes was also given in the above works.
In this work the possibility of production of these two medically relevant radionuclides at a cyclotron was investigated using proton beams on samarium targets. From a detailed study of the literature we recognized already earlier that activation data on rare earth elements are largely missing. It is however well known that many rare earth radionuclides are effectively used or are under investigation for diagnostic nuclear medicine (PET=Positron Emission Tomography) and, in a larger proportion, for therapeutic (radiopharmaceuticals and brachytherapy) purposes. Apart from the medical relevance study of samarium activation has its importance in space and nuclear technology, where the activation via high and medium energy primer and secondary charged particles is unavoidable.
We have measured the activation cross sections of many long-lived radio-products induced in samarium targets, among them the $^{145}$Sm and $^{153}$Sm isotopes. No earlier experimental cross sections were found in the literature as only a few thick target yield data were measured by Dmitriev et al. \cite{7}.
The results are presented and discussed in comparison with theoretical predictions obtained by the nuclear reaction model codes TALYS, EMPIRE and ALICE.

\section{Experiment and data evaluation}
\label{2}
The experimental techniques and data analysis were similar or identical as described by us in recent publications \cite{3, 4}. Here we present shortly the most important factors, and details specific for this experiment in tabular form (Table 1), and shortly in text form.
Taking into account the wide energy range studied, the foil stacks (Sm, monitors and degraders) were irradiated at two incident energies. A first stack (series 1), was irradiated at the Cyclone 90 cyclotron of the Université Catholique in Louvain la Neuve (LLN) with a 65 MeV incident energy proton beam and a second stack (series 2) was irradiated at the CGR 560 cyclotron of the Vrije Universiteit Brussel VUB with a 36 MeV incident energy proton beam.
The activity produced in the targets and monitor foils were measured non-destructively (without chemical separation) using high-resolution (FWHM = 1.6 keV at 1332 keV) HPGe gamma-ray spectrometers. Four series of measurements were done for both irradiations at 10 and 5 cm source -detector distances from 10 min (first spectra series) up to 2 days (for the last spectra series) after end of bombardment (EOB).  The evaluation of the gamma-ray spectra were made by both a commercial and an interactive peak fitting code.
The cross sections were calculated from the well-known activation formula with measured activity, particle flux and number of target nuclei as input parameters. Some of the radionuclides formed are the result of cumulative processes as decay of metastable states or parent nuclides contribute to the production process. Naturally occurring samarium is composed of 7 stable isotopes ($^{144}$Sm -3.07 \%, $^{147}$Sm -14.99 \%, $^{148}$Sm -11.24 \%, $^{149}$Sm -13.82 \%, $^{150}$Sm -7.38 \%, $^{152}$Sm -26.75 \% and $^{154}$Sm -22.75 \%), therefore in most cases so called elemental cross sections were deduced, supposing the Sm to be monoisotopic with total number of target atoms being the sum of all stable isotopes.
The decay data were taken from the online database NuDat2 \cite{8} and the Q-values of the contributing reactions through the Q-value calculator \cite{9}.
Effective beam energy and the energy scale were determined primary by calculation from incident energy, target thickness and stopping power and finally corrected on the basis  of the excitation functions of the $^{24}$Al(p,x)$^{22,24}$Na and $^{nat}$Ti(p,x)$^{48}$V monitor reactions simultaneously re-measured over the whole energy range. For estimation of the uncertainty of the median energy in the target samples and in the monitor foils, the cumulative errors influencing the calculated energy (incident proton energy, thickness of the foils, beam straggling) have been taken into account. The uncertainty of the energy is in the $\pm$ 0.5 - 1.5 MeV range, increasing towards the end of stack. 
The beam intensity (the number of the incident particles) was obtained preliminary through measuring the charge collected in a short Faraday cup and finally on the basis of the monitor reactions. The uncertainties of the cross sections were estimated in the standard way, the independent relative errors of the linearly contributed processes were summed quadratically \cite{10}. Figure 1 shows the re-measured excitation functions of the monitor reactions in comparison with the IAEA (International Atomic Energy Agency) recommended cross section data after correction of the incident energy and the beam current. 
It should be mentioned that in a few cases we could not find independent gamma lines to assess the produced activity of the investigated radioproducts. In these cases the contributions of the overlapping gamma lines from the decay of the other radioproducts were subtracted.
 
\begin{figure}
\includegraphics[scale=0.3]{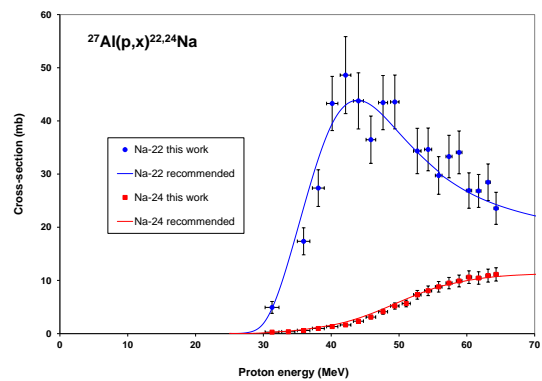}
\caption{The re-measured cross sections of the used monitor reactions in comparison with the recommended data}
\end{figure}

\begin{table*}[t]
\tiny
\caption{Main parameters of the experiment and the methods of data evaluations}
\centering
\begin{center}
\begin{tabular}{|p{1.1in}|p{1.1in}|p{1.1in}|p{1.1in}|p{1.1in}|} \hline 
\multicolumn{3}{|p{3.3in}|}{\textbf{Experiment}} & \multicolumn{2}{|p{2.2in}|}{\textbf{Data evaluation}} \\ \hline 
 & Exp. 1 & Exp.2 & & \\
\hline
Incident particle & Proton & Proton & Gamma spectra evaluation & Genie 
2000, \cite{11}, Forgamma \cite{12} \\
\hline
Method & Stacked foil & Stacked foil & Determination of beam intensity & 
Faraday cup (preliminary)\newline Fitted monitor reaction (final) \cite{13} \\
\hline
Target stack and thicknesses & Al-Sm-Al-Os-Ni-Al block\newline Repeated 20 times\newline
$^{nat}$Al foil, 98 $\mu$m\newline $^{nat}$Sm foil, 100 $\mu$m \newline $^{nat}$Os 
sedimented layer 0.54-2.59 $\mu$m \newline$^{nat}$Ni foil, 25 $\mu$m & 
Ti-Sm-Ti-Al-Mo-Al block\newline Repeated 18 times \newline $^{nat}$Ti foil, 10.9  $\mu$m\newline $^{
nat}$Sm foil, 26.85  $\mu$m\newline $^{nat}$Al foil, 151.12 $\mu$m\newline $^{nat}$Mo 
foil, 11.8 $\mu$m & Decay data & NUDAT 2.6 \cite{8} \\
\hline
Number of $^{nat}$Sm target foils & 20 & 18 & Reaction Q-values & 
Q-value calculator \cite{9} \\
\hline
Accelerator & Cyclone 90 cyclotron Universit\'e Catholique Louvain la 
Neuve (LLN) & CGR 560 cyclotron Vrije Universiteit Brussels(VUB) & 
Determination of beam energy & Andersen (preliminary) \cite{14}\newline Fitted 
monitor reaction (final) \cite{15,16} \\
\hline
Primary energy & 65 MeV & 36 MeV & Uncertainty of energy & Cumulative 
effects of possible uncertainties \\
\hline
Irradiation time & 60 min & 60 min & Cross sections & Elemental cross 
section \\
\hline
Beam current & 20 nA & 83 nA & Uncertainty of cross sections & Sum in 
quadrature of all individual contribution $[$10$]$ \\
\hline
Monitor reaction, $[$recommended values$]$ & $^{nat}$Al(p,x)$^{
22,24}$Na reaction $[$16$]$ & $^{nat}$Ti(p,x)$^{48}$V reaction 
$[$16$]$ & Yield & Physical yield \cite{17} \\
\hline
Monitor target and thickness & $^{27}$Al, 98 $\mu$m & $^{nat}$Ti, 10.9 
 $\mu$m & Theory & ALICE-IPPE \cite{18}, EMPIRE \cite{19}, TALYS (TENDL-2013) 
\cite{20} \\
\hline
detector & HPGe & HPGe & & \\
\hline
$\gamma$-spectra measurements & 4 series & 4 series & & \\
\hline
Cooling times & 7.2-11.6 h\newline 43.7-72.3 h\newline 119-367 h\newline 2142-2569 h & 3.3-9.1 
h\newline 49.9-70.9 h\newline 577-722 h\newline 1951-2237 h & & \\
\hline
\end{tabular}

\end{center}

\end{table*}

\begin{table*}[t]
\tiny
\caption{ Decay characteristics of the investigated activation products and Q-values of contributing reactions}
\centering
\begin{center}
\begin{tabular}{|p{1.1in}|p{0.5in}|p{0.6in}|p{0.5in}|p{0.7in}|p{0.7in}|}
\hline
\textbf{Nuclide\newline Decay path} & \textbf{Half-life} & \textbf{E$_{\gamma
}$(keV)} & \textbf{I$_{\gamma}$(\%)} & \textbf{Contributing reaction
} & \textbf{GS-GS\newline Q-value(keV)} \\
\hline
\textbf{$^{154m}$Eu}\newline IT: 100\%\newline 145.3 keV & 46.0 min & 68.17\newline 100.88 
& 37\newline 27 & $^{154}$Sm(p,n) & -1499.69 \\
\hline
\textbf{$^{154g}$Eu}\newline $\beta^{-}$: 99.982 & 8.601 a & 
123.0706\newline 723.3014\newline 873.1834\newline 996.29\newline 1004.76\newline  1274.429 & 
40.4\newline 20.06\newline 12.08\newline 10.48\newline 18.01\newline 34.8 & $^{154}$Sm(p,n) & -1499.69 \\
\hline
\textbf{$^{152m2}$Eu}\newline IT\newline 147.81 keV & 96 min & 89.849 & 69.7 & 
$^{152}$Sm(p,n)\newline $^{154}$Sm(p,3n) & -2656.944\newline -16492.15 \\
\hline
\textbf{$^{152m1}$Eu}\newline $\beta^{-}$: 73 \% $\varepsilon$: 27\%\newline 45.6 keV & 9.3116 h & 121.77\newline 344.29\newline 841.63\newline 963.38 & 
7.0\newline 2.4\newline  14.2\newline 11.6 & $^{152}$Sm(p,n)\newline $^{154}$Sm(p,3n) & 
-2656.944\newline -16492.15 \\
\hline
\textbf{$^{152g}$Eu}\newline $\beta^{-}$: 27.92 \% $\varepsilon$: 72.08\textit{ }\% 
& 13.517 a & 
121.7817\newline 344.2785\newline 778.9045\newline 244.6974\newline 964.057\newline 1085.837\newline 1112.076\newline 1408.013 & 
28.53\newline 26.59\newline 12.93\newline  7.55\newline 14.65\newline 10.1\newline 13.67\newline  20.87 & $^{152}$Sm(p,n)\newline $^{154}$
Sm(p,3n) & -2656.944\newline -16492.15 \\
\hline
\textbf{$^{150m}$Eu}\newline $\varepsilon$: 11 \textit{ }\% $\beta^{-}$: 89 \%\newline 42.7 keV 
& 12.8 h & 333.9\newline  406.5 &  4.0\newline  2.8 & $^{150}$Sm(p,n)\newline $^{152}$
Sm(p,3n)\newline $^{154}$Sm(p,5n) & -3040.99\newline -16895.14\newline -30730.34 \\
\hline
\textbf{$^{150g}$Eu}\newline $\varepsilon$: 100 \% & 36.9 a & 
333.971\newline 439.401\newline 584.274\newline 1049.043 & 95.16\newline 79.6\newline  52.1\newline  5.33 & $^{150}$
Sm(p,n)\newline $^{152}$Sm(p,3n)\newline $^{154}$Sm(p,5n) & 
-3040.99\newline -16895.14\newline -30730.34 \\
\hline
\textbf{$^{149}$Eu}\newline $\varepsilon$: 100 \%  & 93.1 d & 277.089\newline 327.526 & 
3.56\newline 4.03 & $^{149}$Sm(p,n)\newline $^{150}$Sm(p,2n)\newline $^{152}$Sm(p,4n)\newline 
$^{154}$Sm(p,6n) & -1476.96\newline -9463.65\newline -23317.8\newline -37152.99 \\
\hline
\textbf{$^{148}$Eu}\newline $\varepsilon$: 100 \%  & 54.5 d & 
414.028\newline 414.057\newline 550.284\newline  553.231\newline 553.260\newline 571.962\newline 611.293\newline 629.987\newline 725.673 & 
10.3\newline  10.1\newline  99\newline 12.9\newline 5.0\newline  9.6\newline  20.5\newline 71.9\newline 12.7 & $^{148}$Sm(p,n)\newline $^{149}$
Sm(p,2n)\newline $^{150}$Sm(p,3n)\newline $^{152}$Sm(p,5n)\newline $^{154}$Sm(p,7n) & 
-3819.0\newline -9689.4\newline -17676.1\newline -31530.2\newline -45365.4 \\
\hline
\textbf{$^{147}$Eu}\newline $\varepsilon$: 99.99\%  $\alpha$: 0.0022 \% & 24.1 d & 
 121.220\newline 197.299\newline 677.516\newline 1077.043 &  21.2\newline 24.4\newline 9.0\newline  5.69 & $^{147}$Sm(p,n)\newline 
$^{148}$Sm(p,2n)\newline $^{149}$Sm(p,3n)\newline $^{150}$Sm(p,4n)\newline $^{152}$
Sm(p,6n)\newline $^{154}$Sm(p,8n) & 
-2503.96\newline -10645.33\newline -16515.69\newline -24502.37\newline -38356.52\newline -52191.72 \\
\hline
\textbf{$^{146}$Eu}\newline $\varepsilon$: 100 \%  & 4.59 d & 
632.89\newline 633.083\newline  634.137\newline 747.159 & 1.28\newline 35.9\newline 45.0\newline 99 & $^{147}$Sm(p,2n)\newline $^{
148}$Sm(p,3n)\newline $^{149}$Sm(p,4n)\newline $^{150}$Sm(p,5n)\newline $^{152}$
Sm(p,7n)\newline $^{154}$Sm(p,9n) & 
-11002.26\newline -19143.63\newline -25013.98\newline -33000.67\newline -46854.82\newline -60690.02 \\
\hline
\textbf{$^{145}$Eu}\newline $\varepsilon$: 100 \%  & 5.93 d &  653.512\newline 893.73\newline 1658.53 
& 15.0\newline 66\newline 14.9 & $^{147}$Sm(p,3n)\newline $^{148}$Sm(p,4n)\newline $^{149}$
Sm(p,5n)\newline $^{150}$Sm(p,6n)\newline $^{152}$Sm(p,8n)\newline $^{154}$Sm(p,10n) & 
-18199.45\newline -26340.83\newline -32211.18\newline -40197.87\newline -54052.02\newline -63004.8 \\
\hline
\textbf{$^{153}$Sm}\newline $\beta^{-}$: 100 \% & 46.50 h & 103.18012 & 
29.25 & $^{154}$Sm(p,pn)\newline $^{153}$Pm decay & -7966.79\newline -9095.91 \\
\hline
\textbf{$^{145}$Sm}\newline $\varepsilon$: 100 \%  & 340 d & 61.2265 & 12.15 & $^{147
}$Sm(p,p2n)\newline $^{148}$Sm(p,p3n)\newline $^{149}$Sm(p,p4n)\newline $^{150}$
Sm(p,p5n)\newline $^{152}$Sm(p,p7n)\newline $^{154}$Sm(p,p9n)\newline $^{145}$Eu decay 
& -14757.41\newline -22898.8\newline -28769.15\newline -36755.84\newline -50609.98\newline -64445.16\newline -18199.45 \\
\hline
\textbf{$^{151}$Pm}\newline $\beta^{-}$: 100 \%  & 28.40 h & 
240.09\newline 275.21\newline 340.08\newline 445.68\newline  717.72 & 3.8\newline 6.8\newline 22.5\newline  4.0\newline 4.1 & $^{152}$
Sm(p,2p)\newline $^{154}$Sm(p,2p2n) & -8665.55\newline -22500.74 \\
\hline
\textbf{$^{150}$Pm}\newline $\beta^{-}$: 100 \%  & 2.68 h & 
333.92\newline  406.51\newline 831.85\newline 876.41\newline 1165.77\newline 1323.28  &  68\newline  5.6\newline  11.9\newline 7.3\newline 15.8\newline 2.8 & $^{
152}$Sm(p,2pn)\newline $^{154}$Sm(p,2p3n) & -16525.8\newline -30361.0 \\
\hline
\textbf{$^{149}$Pm}\newline $\beta^{-}$: 100 \%  & 53.08 h & 285.95  & 3.1 
& $^{150}$Sm(p,2p)\newline $^{152}$Sm(p,2p2n)\newline $^{154}$Sm(p,2p4n) & 
-8275.76\newline -22129.9\newline -35965.09 \\
\hline
\end{tabular}
\end{center}
\end{table*}

\setcounter{table}{1}

\begin{table*}[t]
\tiny
\caption{continued}
\centering
\begin{center}
\begin{tabular}{|p{1.1in}|p{0.5in}|p{0.6in}|p{0.5in}|p{0.7in}|p{0.7in}|}
\hline
\textbf{Nuclide\newline Decay path} & \textbf{Half-life} & \textbf{E$_{\gamma
}$(keV)} & \textbf{I$_{\gamma}$(\%)} & \textbf{Contributing reaction
} & \textbf{GS-GS\newline Q-value(keV)} \\
\hline
\textbf{$^{148m}$Pm}\newline $\beta^{-}$: 95.8 \% IT: 4.2 \%\newline 137.93 keV & 
41.29 d & 288.11\newline 414.07\newline 432.745\newline 550.284\newline 629.987\newline 725.673\newline 915.331\newline 1013.808\newline 
& 12.56\newline 18.66\newline 5.35\newline 94.9\newline  89.0\newline 32.8\newline 17.17\newline 20.3 & $^{149}$Sm(p,2p)\newline $^{150}$
Sm(p,2pn)\newline $^{152}$Sm(p,2p3n)\newline $^{154}$Sm(p,2p5n) & 
-7558.99\newline -15545.68\newline -29399.83\newline -43235.02 \\
\hline
\textbf{$^{148g}$Pm}\newline $\beta^{-}$: 100 \% & 5.368 d & 
 550.27\newline  914.85\newline 1465.12 & 22.0\newline 11.5\newline 22.2 & $^{149}$Sm(p,2p)\newline $^{150}$
Sm(p,2pn)\newline $^{152}$Sm(p,2p3n)\newline $^{154}$Sm(p,2p5n) & 
-7558.99\newline -15545.68\newline -29399.83\newline -43235.02 \\
\hline
\textbf{$^{147}$Pm}\newline $\beta^{-}$: 100 \% & 2.6234 a & 121.220 & 
0.00285 & & \\
\hline
\textbf{$^{146}$Pm}\newline $\varepsilon$: 66.0\% ?$^{-}$: 34.0
\% &  5.53 a & 453.88\newline 633.25\newline 735.9\newline 747.24 & 65.\newline  2.15\newline  22.5\newline 34 & 
$^{147}$Sm(p,2p)\newline $^{148}$Sm(p,2pn)\newline $^{149}$Sm(p,2p2n)\newline $^{150
}$Sm(p,2p3n)\newline $^{152}$Sm(p,2p5n)\newline $^{154}$Sm(p,2p7n) & 
-7100.77\newline -15242.15\newline -21112.5\newline -29099.19\newline -42953.34\newline -56788.53 \\
\hline
\textbf{$^{145}$Pm}\newline $\varepsilon$: 100 \% & 17.7 a & 67.2\newline  72.4 & 0.68\newline 2.20 & $^{
147}$Sm(p,2pn)\newline $^{148}$Sm(p,2p2n)\newline $^{149}$Sm(p,2p3n)\newline $^{150}$
Sm(p,2p4n)\newline $^{152}$Sm(p,2p6n)\newline $^{154}$Sm(p,2p8n) & 
-13358.98\newline -21500.35\newline -27370.7\newline -35357.39\newline -49211.54\newline -63046.73 \\
\hline
\textbf{$^{144}$Pm}\newline $\varepsilon$: 100 \% & 363 d & 476.78\newline 618.01\newline 696.49 & 
 43.8\newline  98\newline 99.49 & $^{147}$Sm(p,2p2n)\newline $^{148}$Sm(p,2p3n)\newline $^{149}$
Sm(p,2p4n)\newline $^{150}$Sm(p,2p5n)\newline $^{152}$Sm(p,2p7n)\newline $^{154}$
Sm(p,2p9n) & -21281.69\newline -29423.06\newline -35293.41\newline -43280.1\newline -57134.24\newline -70969.43 \\
\hline
\textbf{$^{143}$Pm}\newline $\varepsilon$: 100 \% & 265 d & 741.98 & 38.5 & $^{144
}$Sm(p,2p)\newline $^{147}$Sm(p,2p3n)\newline $^{148}$Sm(p,2p4n)\newline $^{149}$
Sm(p,2p5n)\newline $^{150}$Sm(p,2p6n)\newline $^{152}$Sm(p,2p8n)\newline $^{154}$
Sm(p,2p10n)\newline $^{143}$Sm decay & 
-6293.94\newline -27808.46\newline -35949.84\newline -41820.19\newline -49806.87\newline -63661.01\newline -71150.82\newline -8295.49 \\
\hline
\textbf{$^{141}$Nd}\newline $\varepsilon$: 100 \% & 2.49 h & 1126.91\newline 1292.64 & 0.80\newline 0.46 
& $^{144}$Sm(p,3pn)\newline $^{147}$Sm(p,3p4n)\newline $^{148}$Sm(p,3p5n)\newline $^{
149}$Sm(p,3p6n)\newline $^{141}$Pm decay & 
-20421.3\newline -41935.82\newline -50077.14\newline -55947.53\newline 3421.5 \\
\hline
\end{tabular}
\end{center}

\begin{flushleft}
\tiny{\noindent Isotopic abundance: $^{144}$Sm-3.1 \%-stable, $^{147}$Sm -15.0 \%-1.06*1011 a, $^{148}$Sm -11.3 \%-7*1015 a, $^{149}$Sm -13.8 \%-7*1015 a, $^{150}$Sm -7.4 \%- stable, $^{152}$Sm -26.7- stable, $^{154}$Sm -22.7 \%- stable\newline
When complex particles are emitted instead of individual protons and neutrons the Q-values have to be decreased by the respective binding energies of the compound particles: np-d, +2.2 MeV; 2np-t, +8.48 MeV; 2p2n-$\alpha$, 28.30 MeV\newline
Decrease Q-values for isomeric states with level energy of the isomer}
\end{flushleft}

\end{table*}

\section{Model calculations}
\label{3}
The ALICE-IPPE \cite{18} and EMPIRE \cite{19} codes were used to analyse the present experimental results. The parameters for the optical model, level densities and pre-equilibrium contributions were taken as described in \cite{15}. The cross sections for isomers in case of ALICE code were obtained by using the isomeric ratios calculated with EMPIRE.  The theoretical data from the TENDL-2013 library (based on the TALYS 1.4 code \cite{21} was used for a comparison too.

\section{Results and discussion}
\label{4}

\subsection{Cross sections}
\label{4.1}

The cross sections for all studied $^{nat}$Sm(p,x) reactions are shown in Figures 2 – 14 and the numerical values are collected in Tables 3-4. The reactions contributing to the production of a given activation product and their Q-values are given in Table 2. The radioproducts of Eu are produced via (p,xn) reactions, the samarium and promethium  isotopes  directly  by (p,pxn) and (p,2pxn) reactions and by EC/$\beta^+$ and $\beta^-$decay of parent radioisotopes. As the results of the two irradiations slightly differ in the overlapping energy range, the excitation functions are represented separately. The descriptions of the nuclear parameters of the investigated reactions have been taken from our recent publications \cite{3, 4}. Practically the same radionuclides were detected (except $^{154}$Eu and $^{141}$Nd). The decay data of the discussed radioproducts presented in Table 1 are also repeated for the easier reading. 

\subsubsection{Production of radioisotopes of europium}
\label{4.1.1}

Production cross sections of $^{154g}$Eu 

The production cross sections of the $^{154}$Eu ground state (GS, E(level) = 0.0 keV,  $J^\pi = 3^-$, $T_{1/2}$ = 8.601 a, $\beta^-$: 99.982 \%, $\varepsilon$: 0.018 \%) contain the direct production and the  contribution of complete decay of the short lived  isomeric state  (MS, E(level) = 145.3 keV,  $J^\pi = 8^-$, $T_{1/2}$ = 46.0 min,  IT:100 \%) \cite{3, 4}. The relative long cooling time of the first spectra (and the low cross sections) not allows detecting the decay of the high spin isomeric state.  Due to the long half-life, and the low beam intensity $^{154g}$Eu was detected with poor statistics only in some of the long measurement (Fig. 2). The $^{154g}$Eu and the short half-life isomeric state is produced through the $^{154}$Sm(p,n) reaction. 

\begin{figure}
\includegraphics[scale=0.3]{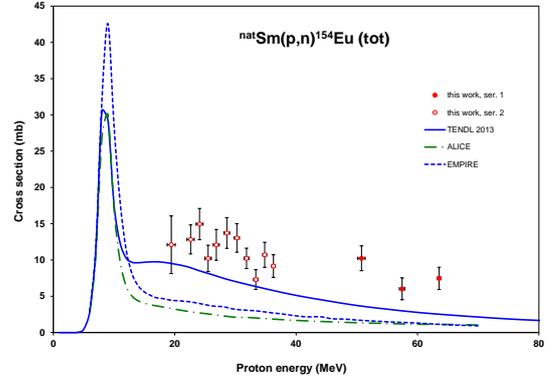}
\caption{Excitation functions of the $^{nat}$Sm(p,xn)$^{154}$Eu reaction in comparison with results from model calculations}
\end{figure}

Production cross sections of different levels of $^{152}$Eu  

The radionuclide $^{152}$Eu has, apart from its long-lived ground state (GS, E(level)= 0.0  keV, $T_{1/2}$= 13.517 a, $J^\pi = 3^-$, $\beta^-$: 27.9  \%,  $\varepsilon$: 72.1  \%), two longer-lived isomeric states with complex decay scheme. The high spin, rather short-lived, isomer $^{152m2}$Eu (MS2, E(level) =147.81 keV, $J^\pi = 8^-$, $T_{1/2}$= 96 min, IT  100 \% ) decays totally by isomeric transition to the ground state, without population of the lower energy $^{152m1}$Eu level. 
This latter metastable state with zero spin (MS1, E(level) = 45.5998 keV, $J^\pi = 0^-$, $T_{1/2}$=  9.3116 h, $\beta^-$: 73  \%, $\varepsilon$: 27  \%) decays by $\beta^-$ decay to stable $^{152}$Gd and does not populate the ground state. Independent gamma lines are emitted in the decay of the different states so that production cross sections for all three levels can be determined \cite{3, 4}.

Production cross sections of $^{152m2}$Eu 

The results for the high spin isomer (MS2, E(level) =147.81 keV, $J^\pi = 8^-$, $T_{1/2}$ = 96 min, IT:100 \% ) are shown in Fig. 3. The theory significantly overestimates the concordant values of the two experiments.

\begin{figure}
\includegraphics[scale=0.3]{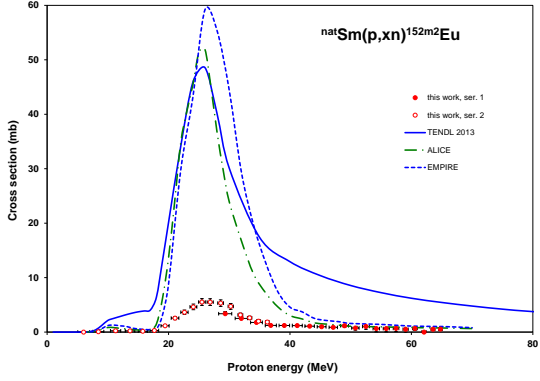}
\caption{Excitation functions of the $^{nat}$Sm(p,xn)$^{152m2}$Eu reaction in comparison with results from model calculations}
\end{figure}

Production cross sections of $^{152m1}$Eu

The results for the low spin isomer (MS1, E(level) = 45.5998 keV, $J^\pi = 0^-$, $T_{1/2}$ =  9.3116 h, $\beta^-$: 72  \%, $\varepsilon$: 27  \%) are shown in Fig. 4. Experimentally, the contribution of the reactions on $^{152}$Sm and $^{154}$Sm can be distinguished in the results of the low energy experiment.  The agreement with the theoretical predictions is acceptable.

\begin{figure}
\includegraphics[scale=0.3]{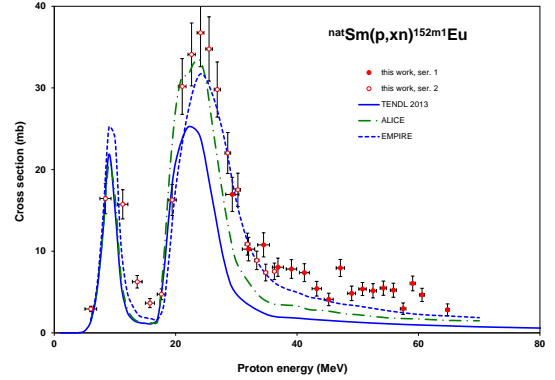}
\caption{Excitation functions of the $^{nat}$Sm(p,xn)$^{152m1}$Eu reaction in comparison with results from model calculations}
\end{figure}

Production cross sections of $^{152g}$Eu

The measured cross sections for production of the ground state (GS, E(level)= 0.0  keV, $T_{1/2}$=13.537 a, $J^\pi = 3^-$, $\beta^-$: 27.9\%, $\varepsilon$: 72.1 \%) are shown in Fig. 5. The uncertainties of the results of the high energy irradiation are high due to the low activity. As the measurement were taken after a long cooling time, in addition to the direct production also the contribution of total decay of  the high spin state $^{152m2}$Eu (IT: 100  \%) is included (m2+).  

\begin{figure}
\includegraphics[scale=0.3]{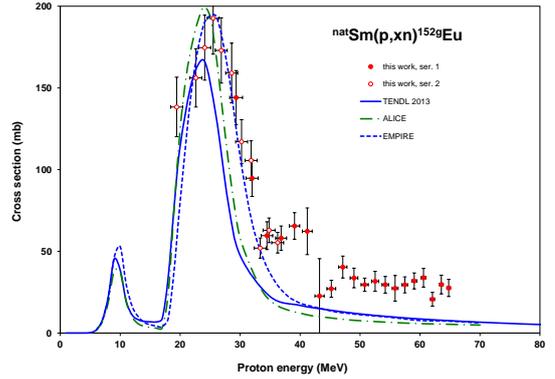}
\caption{Excitation functions of the $^{nat}$Sm(p,xn)$^{152g}$Eu reaction in comparison with results from model calculations}
\end{figure}

Production cross sections of different levels of $^{150}$Eu

The radioisotope $^{150}$Eu has two isomeric states, a short-lived excited and a long-lived ground state. 

Production cross sections of $^{150m}$Eu

The cross section of the-long lived isomeric state of $^{150}$Eu (MS, E(level) = 42.1 keV, $J^\pi = 0^-$, $T_{1/2}$ = 12.8 h , $\varepsilon$: 11  \% , $\beta^-$ 89  \%, IT <5E-8) is shown in Fig. 6.  The agreement of the theoretical and experimental data is acceptable. The experimental data are slightly lower over the whole energy range.

\begin{figure}
\includegraphics[scale=0.3]{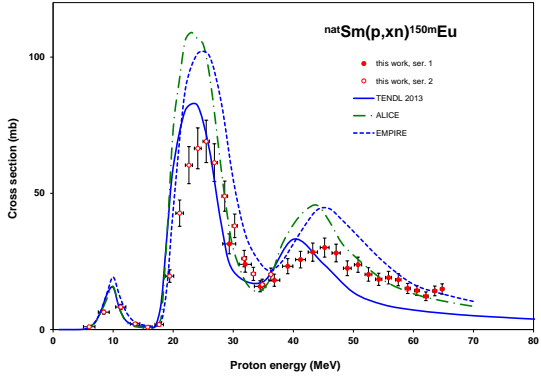}
\caption{Excitation functions of the $^{nat}$Sm(p,xn)$^{150m}$Eu reaction in comparison with results from model calculations}
\end{figure}

Production cross sections of $^{150g}$Eu

Only direct production has to be considered as practically no contribution from the decay of $^{150m}$Eu exists (IT < 10E-8, see previous paragraph).  The independent cross sections for direct production of $^{152g}$Eu (GS, E(level)= 0.0 keV, $J^\pi = 5^-$, $T_{1/2}$ = 36.9 a,  $\varepsilon$: 100 \% ) are compared with the theory in Fig 7.  The theoretical predictions support our data. The data from Gheorghe et al. \cite{22} has no overlap with our results.

\begin{figure}
\includegraphics[scale=0.3]{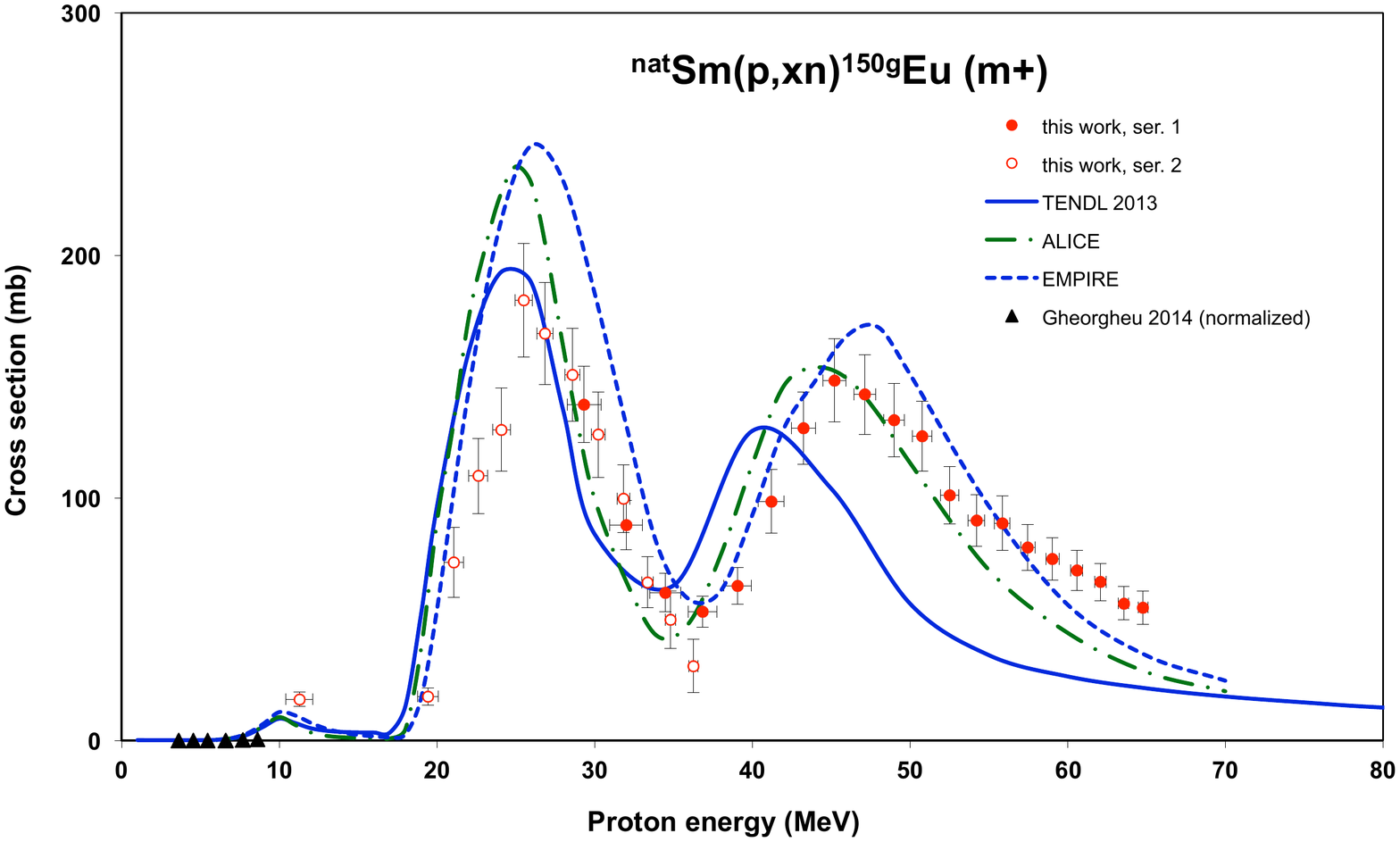}
\caption{Excitation functions of the $^{nat}$Sm(p,xn)$^{150g}$Eu reaction in comparison with results from model calculations}
\end{figure}

Production cross sections of $^{149}$Eu

Only direct (p,xn) reactions on different stable Sm isotopes contribute to the  production of $^{149}$Eu (GS, E(level)=0.0  keV, $J^\pi = 3{5/2}^+$, 	  $T_{1/2}$=93.1 d, $\varepsilon$: 100 \%). The theoretical results are close to our experimental data below 25 MeV (contributions of (p,n) and (p,2n) (Fig.8). From the onset of the (p,4n) reaction on (25 MeV) the theoretical  prediction of TENDL-2013 differs both in shape and in the magnitude from the experiments. The data of Gheorghe et al. \cite{22} show good agreement with our in the overlapping energy range (2 points).

\begin{figure}
\includegraphics[scale=0.3]{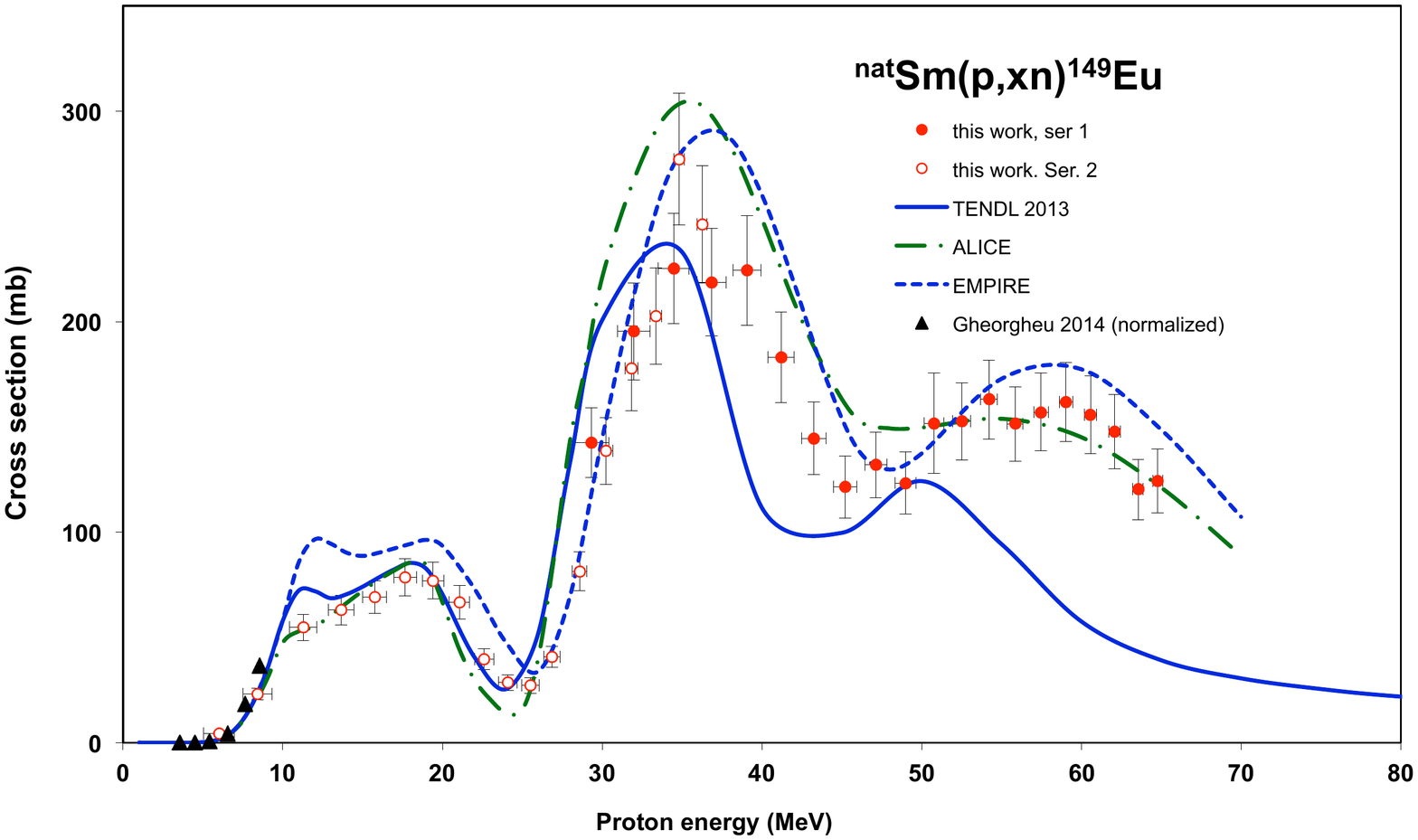}
\caption{Excitation functions of the $^{nat}$Sm(p,xn)$^{149}$Eu reaction in comparison with results from model calculations}
\end{figure}

Production cross sections of $^{148}$Eu

The experimental and theoretical data for cross sections of $^{148}$Eu (GS, E(level)= 0.0 keV, $J^\pi = 5^-$, $T_{1/2}$=54.5 d,  $\varepsilon$: 100 \%, $\alpha$: 9.4E-7 \%) are compared in Fig. 9. In the energy domain considered reactions on four stable Sm isotopes can contribute. The agreement with the theory at the low energy range is acceptable. At high energies the theoretical data of the different model codes differ significantly. The data of Gheorghe et al. \cite{22} are much lower than ours. Their measurements were done on enriched $^{147}$Sm target with less known enrichment ratio and only the (p,$\gamma$) process was measured/published. This obviously shows that the (p,$\gamma$) process does not play significant role in $^{148}$Eu production.

\begin{figure}
\includegraphics[scale=0.3]{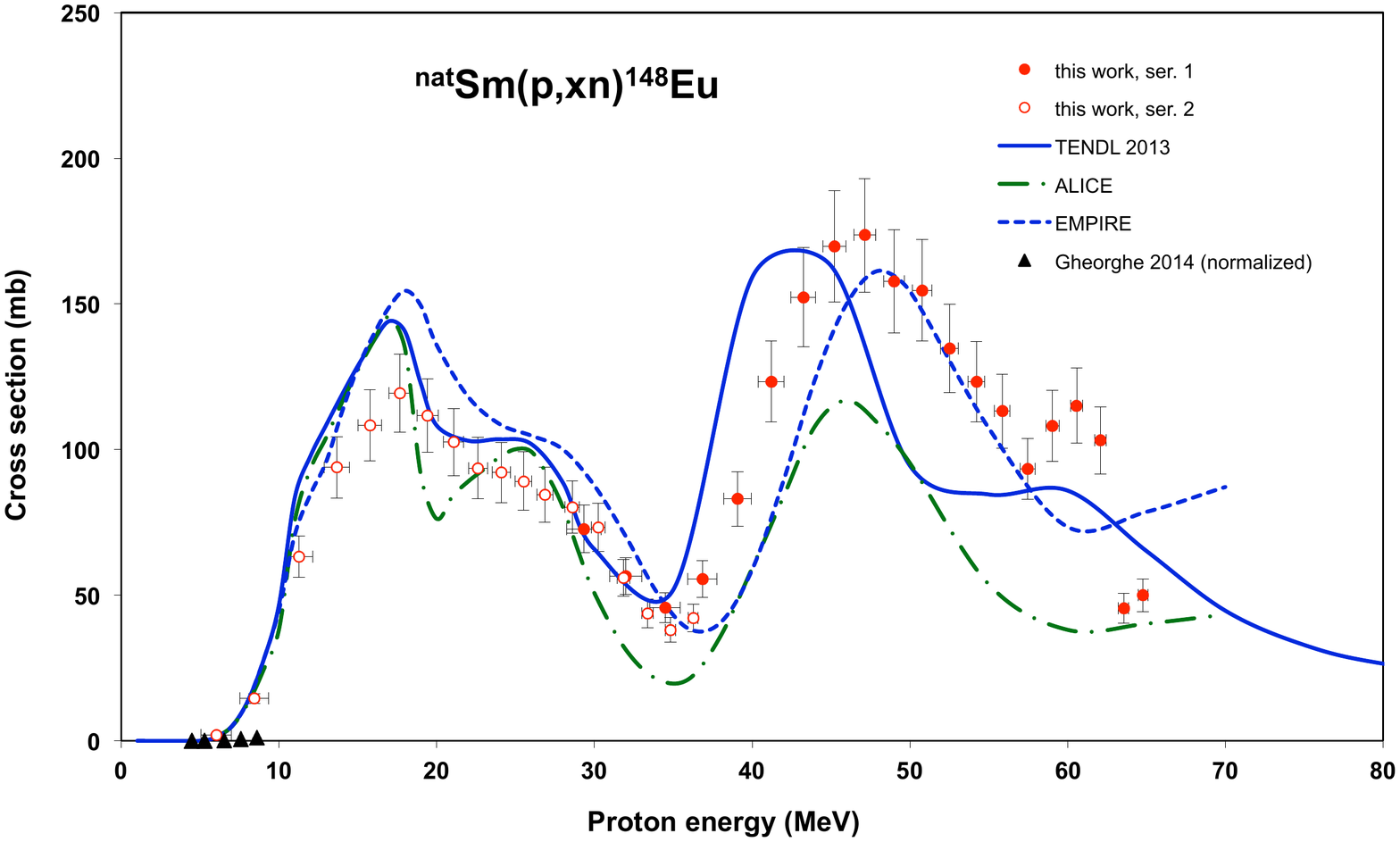}
\caption{Excitation functions of the $^{nat}$Sm(p,xn)$^{148}$Eu reaction in comparison with results from model calculations}
\end{figure}

Production cross sections of $^{147}$Eu

There is acceptable agreement in the shape between the experiment and the theory for $^{147}$Eu (GS, E(level) = 0.0 keV,  $J^\pi = {5/2}^+$, $T_{1/2}$= 24.1 d, $\varepsilon$: 99.9978 \%, $\alpha$: 0.0022 \%) cross sections (Fig. 10). The experiment shows no details from the contributions on all stable Sm target isotopes except for the significant effect of $^{154}$Sm(p,8n) above 50 MeV. The data of Gheorghe et al. \cite{22} show excellent agreement with our results. 

\begin{figure}
\includegraphics[scale=0.3]{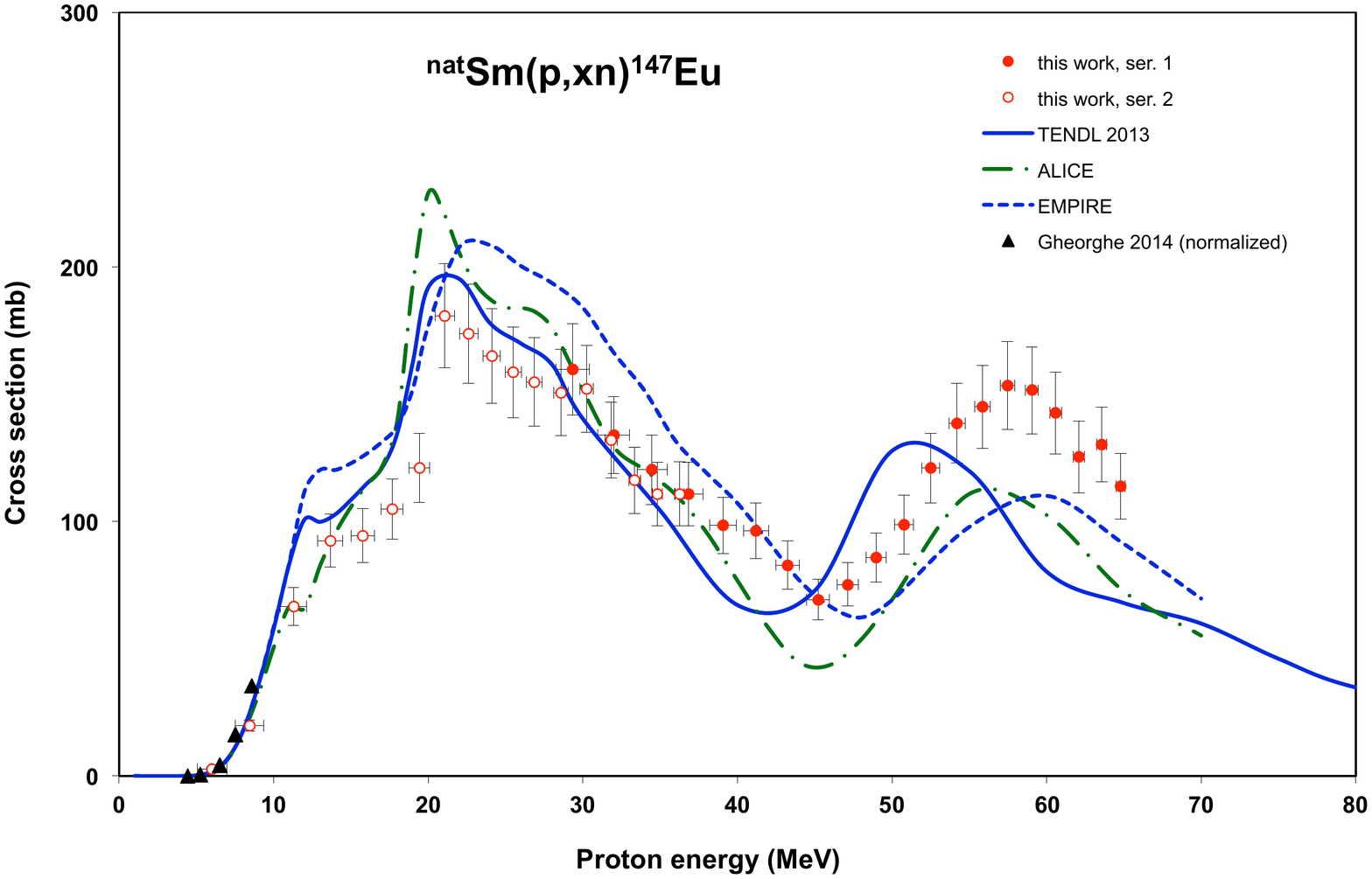}
\caption{Excitation functions of the $^{nat}$Sm(p,xn)$^{147}$Eu reaction in comparison with results from model calculations}
\end{figure}

Production cross sections of $^{146}$Eu

 The experimental results correspond well to the overall shape of the more detailed theoretical descriptions of the excitation function (contribution of the six stable Sm isotopes) for $^{146}$Eu (GS, E(level)= 0.0 keV, $J^\pi = 4^-$, $T_{1/2}$ = 4.59 d, $\varepsilon$: 100 \%).  

\begin{figure}
\includegraphics[scale=0.3]{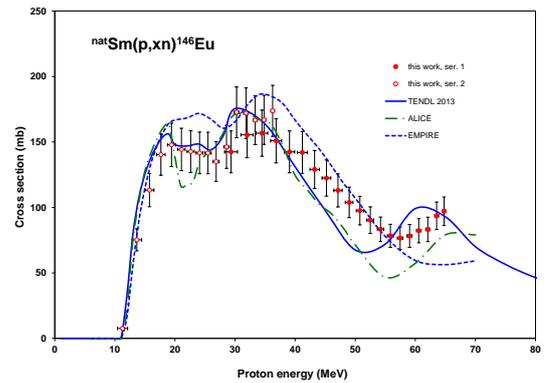}
\caption{Excitation functions of the $^{nat}$Sm(p,xn)$^{146}$Eu reaction in comparison with results from model calculations}
\end{figure}

Production cross sections of $^{145}$Eu

The production cross sections of $^{145}$Eu (GS, E(level)= 0.0 keV,  $J^\pi = {5/2}^+$,  $T_{1/2}$=5.93 d,  $\varepsilon$: 100 \%) are shown in Fig. 12. The agreement with the theory can also be considered acceptable.

\begin{figure}
\includegraphics[scale=0.3]{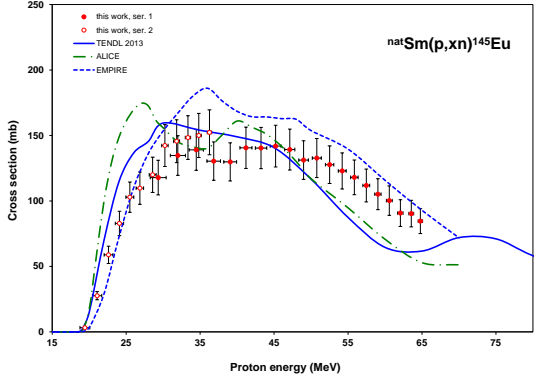}
\caption{Excitation functions of the $^{nat}$Sm(p,xn)$^{145}$Eu reaction in comparison with results from model calculations}
\end{figure}

\subsubsection{Production of radioisotopes of samarium}
\label{4.1.2}

Production cross sections of $^{153}$Sm

The medically relevant radionuclide $^{153}$Sm (GS, E(level) = 0.0 keV, $J^\pi = {3/2}^+$,  $T_{1/2}$ = 46.50 h,  $\beta^-$: 100 \%) is  produced directly through (p,pn) reactions and through the decay of the short-lived 153Pm parent (GS, E(level) =  0.0 keV,  $J^\pi = {5/2}^-$, $T_{1/2}$= 5.25 min,  $\beta^-$: 100 \%.) The presented cross sections are cumulative, obtained from spectra measured after the decay of 153Pm.  The magnitude of the different theoretical data differs significantly (Fig. 13), the TENDL data being nearly two times higher than the values obtained by ALICE-IPPE. Below 25 MeV our experimental values are higher than all theoretical calculations but at higher energies there is an agreement with EMPIRE.

\begin{figure}
\includegraphics[scale=0.3]{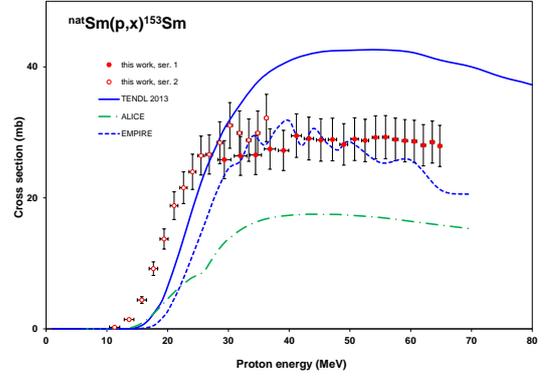}
\caption{Excitation functions of the $^{nat}$Sm(p,x)$^{153}$Sm reaction in comparison with results from model calculations}
\end{figure}

Production cross sections of $^{145}$Sm

The cross sections for cumulative production of $^{145}$Sm (GS, E(level) =  0.0  keV,  $J^\pi = {7/2}$, $T_{1/2}$ = 340 d, $\varepsilon$:100 \%) are shown In Fig. 14. It includes the direct production through (p,pxn) reactions and the contribution from the decay of $^{145}$Eu  (GS,  E(level) =  0.0  keV, $J^\pi = {5/2}^+$, $T_{1/2}$= 5.93 d, $\varepsilon$: 100 \%). The theories predict cross sections close to the experimental data in the investigated energy range

\begin{figure}
\includegraphics[scale=0.3]{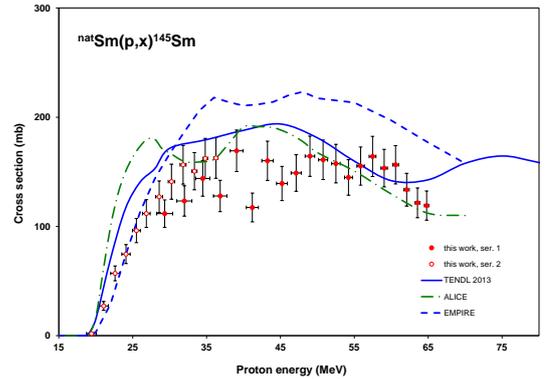}
\caption{Excitation functions of the $^{nat}$Sm(p,x)$^{145}$Sm reaction in comparison with results from model calculations}
\end{figure}

\subsubsection{Production of radioisotopes of promethium}
\label{4.1.3}

Production cross sections of $^{151}$Pm

The cross sections of the $^{151}$Pm (GS, E(level) = 0.0 keV,  $J^\pi = {5/2}^+$, $T_{1/2}$ = 28.40 h,  $\beta^-$: 100 \%) (Fig. 15) are practically independent, direct production cross sections through (p,2pxn) reactions and the cross sections of parent $^{151}$Nd  ($T_{1/2}$ = 12.44 min) obtained by (p,3pxn) are negligible small (as predicted by theory and by the systematics). The sharp peak at low energies and steep rise at high energies that can be seen in the TENDL theoretical data is hence questionable and is not reproduced by the experimental values. Only one data point was measured by Milazzo-Coli et al. in 1974 \cite{23} on enriched $^{154}$Sm with unknown enrichment rate, which might cause the disagreement (we normalized it assuming 100 \% enrichment).

\begin{figure}
\includegraphics[scale=0.3]{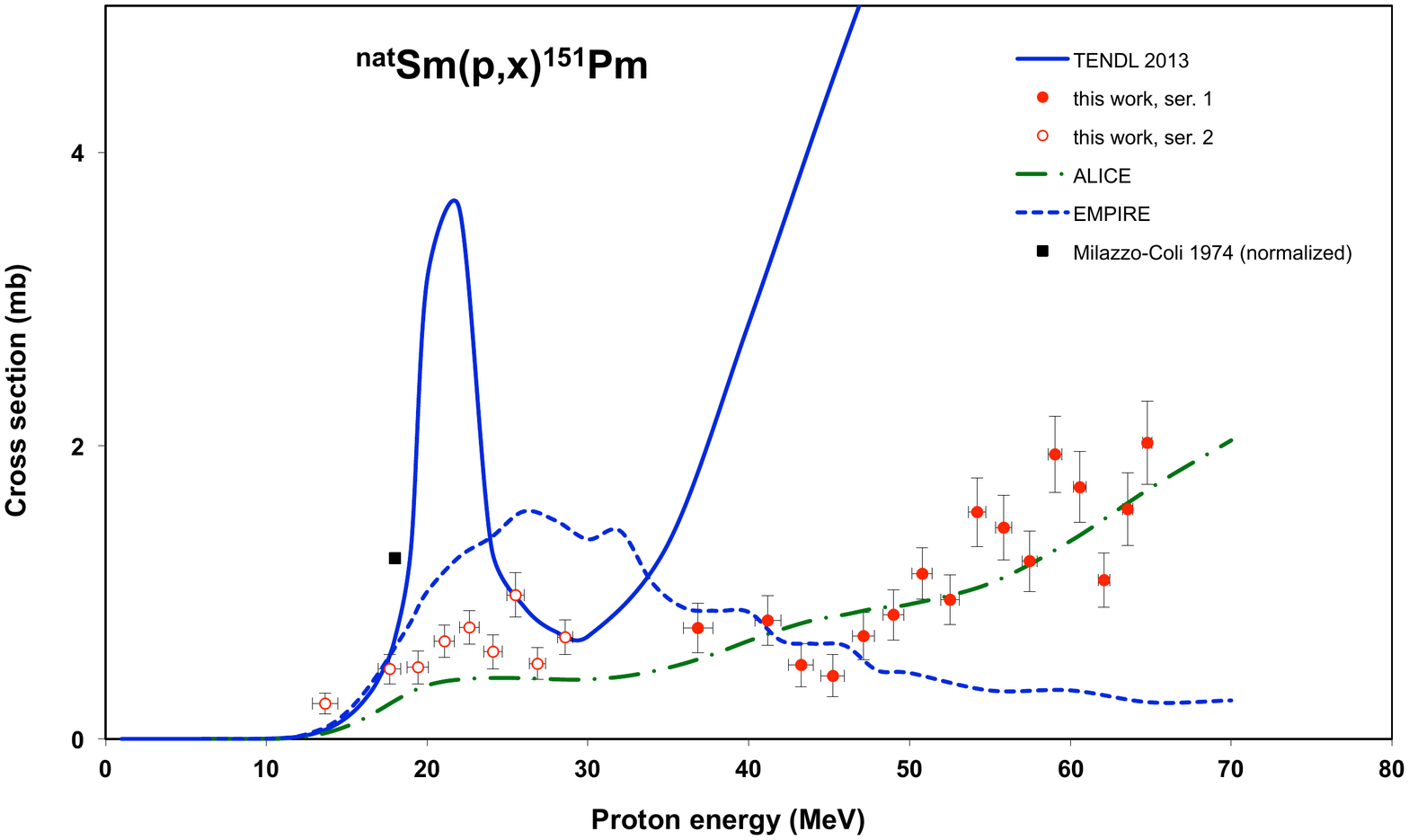}
\caption{Excitation functions of the $^{nat}$Sm(p,x)$^{151}$Pm reaction in comparison with results from model calculations}
\end{figure}

Production cross sections of $^{150}$Pm

 The radionuclide $^{150}$Pm (GS, E(level) = 0.0 keV,  $J^\pi = 1^-$, $T_{1/2}$ = 2.68 h, $\beta^-$: 100 \%) is a closed nuclei from the  point of view of decay of parents. It is hence produced only directly via (p,2pxn) reactions (Fig. 16). The theoretical data of the TENDL code overestimate the experimental results above 45 MeV. The ALICE results are a factor of two lower than the experiment. 

\begin{figure}
\includegraphics[scale=0.3]{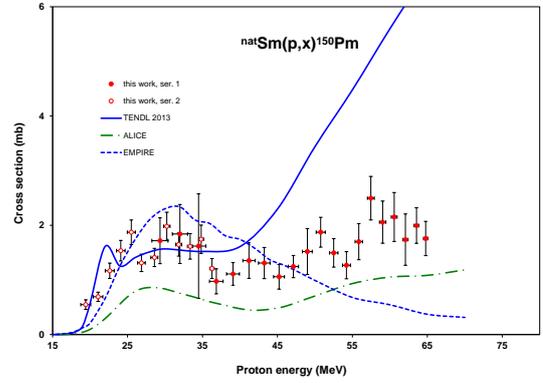}
\caption{Excitation functions of the $^{nat}$Sm(p,x)$^{150}$Pm reaction in comparison with results from model calculations}
\end{figure}

Production cross sections of $^{149}$Pm

Neglecting the small contribution from decay of $^{149}$Nd (GS, E(level) = 0.0 keV, $J^\pi = {5/2}^-$, $T_{1/2}$ = 1.728 h, $\beta^-$: 100 \%) produced by (p,3pxn) reactions with low cross sections, the $^{149}$Pm (GS, E(level) = 0.0 keV, $J^\pi = {7/2}^+$, 53.08 h, $\beta^-$: 100 \%) is produced only directly. We could assess only two experimental points at higher energies (long measurements), where the statistics were sufficient to calculate significant, low cross sections (Fig 17). The decay data were taken from Lundl/LBNL Nuclear Data Base \cite{24}. The results of the three theoretical calculations differ significantly both in shape and in magnitude.

\begin{figure}
\includegraphics[scale=0.3]{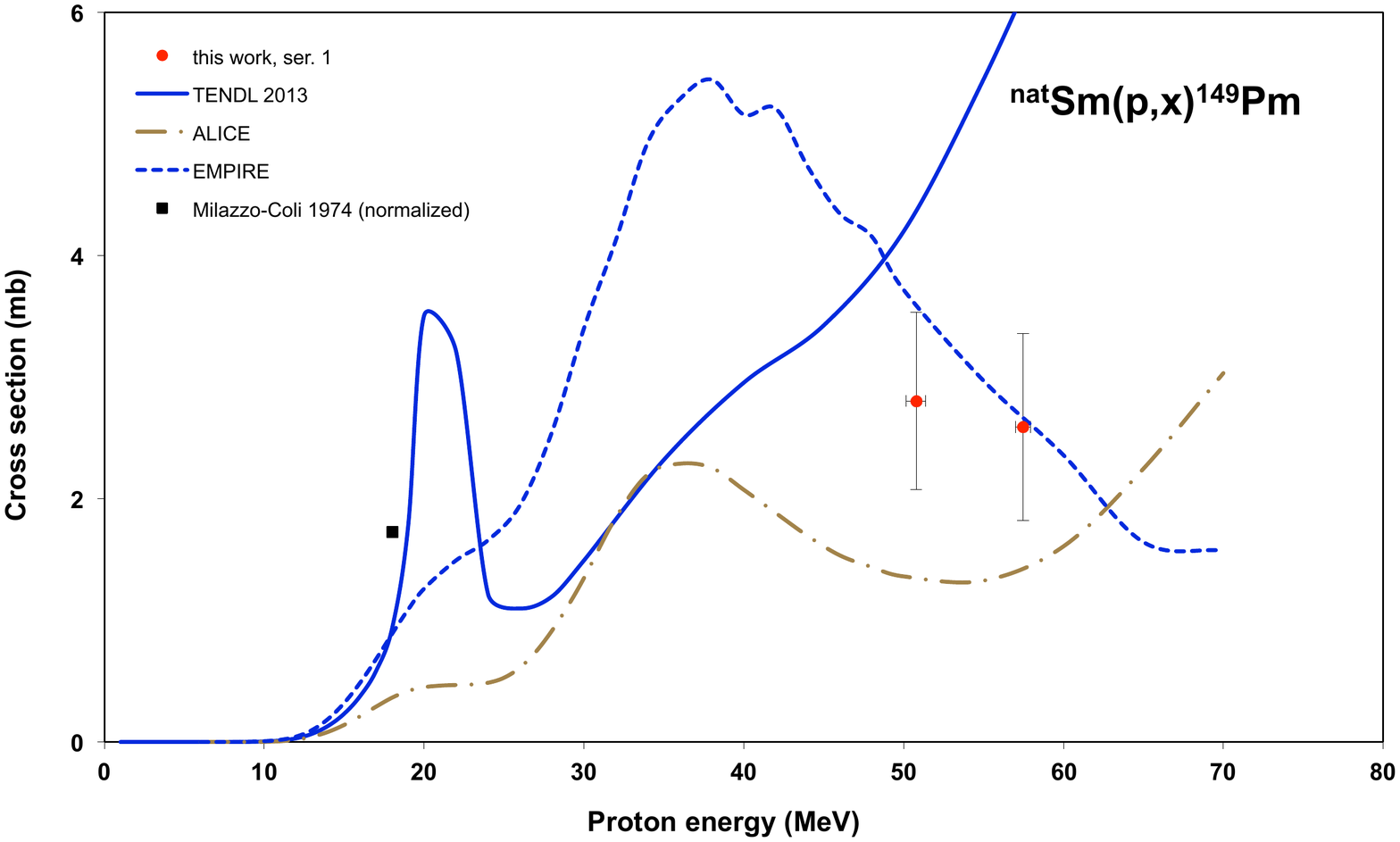}
\caption{Excitation functions of the $^{nat}$Sm(p,x)$^{149}$Pm reaction in comparison with results from model calculations}
\end{figure} 

Production cross sections of different levels of $^{148}$Pm

The radioisotope $^{148}$Pm has two different isomeric states. The metastable state is longer-lived, while the ground-state has somewhat shorter half-life.

Production cross sections of $^{148m}$Pm  

The theoretical cross sections of the directly produced $^{148m}$Pm (MS, E(level) = 137.93 keV,  $J^\pi = 5^-, 6^-$, $T_{1/2}$ = 41.29 d, IT: 4.2 \%, $\beta^-$: 95.8 \%) are shown in Fig 18. The $^{148m}$Pm ($T_{1/2}$ = 41.29 d) and the simultaneously produced $^{148}$Eu ($T_{1/2}$ = 54.5 d) have similar half-lives and both decay to $^{148}$Sm resulting in the fact that practically all gamma-lines of $^{148m}$Pm and $^{148}$Eu are the same. As the production of $^{148}$Eu has significantly higher cross sections, the separation of the gamma intensities can be done only with very large uncertainties. The presented experimental cross section are hence only indicative but are only a factor of two higher than the theoretically predicted values in the overlapping energy range.

\begin{figure}
\includegraphics[scale=0.3]{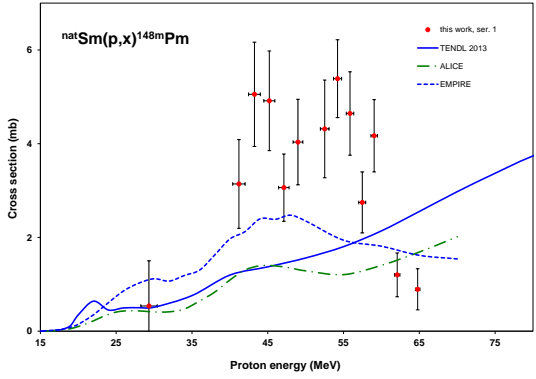}
\caption{Excitation functions of the $^{nat}$Sm(p,x)$^{148m}$Pm reaction in comparison with results from model calculations}
\end{figure}

Production cross sections of $^{148g}$Pm 

The ground state of $^{148}$Pm (GS, E(level) = 0.0 keV,  $J^\pi = 1^-$,  $T_{1/2}$ = 5.368 d, $\beta^-$: 100 \%) is produced directly and from the weak isomeric decay of the longer-lived metastable state (41.29 d,  IT: 4.2 \%) discussed in the previous section. We were not able to identify in any of our spectra the only strong independent gamma-line (1465.12 keV, 22.2 \%) of $^{148g}$Pm.  In principle it is possible to obtain data for direct production of $^{148g}$Pm from an analysis of the time dependence of the count rate for the 550 keV and 915 keV gamma-lines common between the ground state, the isomeric state and $^{148}$Eu, but due to the low cross sections expected and the large inherent uncertainty we decided not to do an unreliable separation. The theoretical results are shown in Fig. 19 and predict an isomeric ratio (M/G) of about 0.5. The experimental ratio is significantly higher. 

\begin{figure}
\includegraphics[scale=0.3]{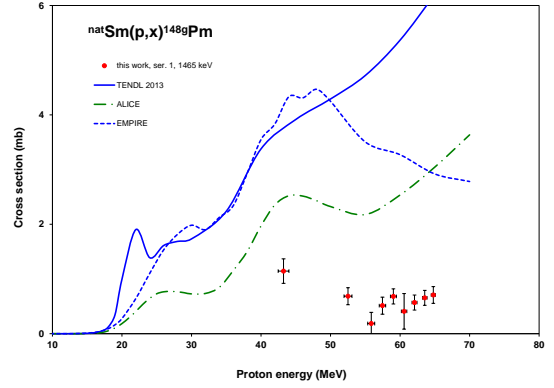}
\caption{Excitation functions of the $^{nat}$Sm(p,x)$^{148g}$Pm reaction in comparison with results from model calculations}
\end{figure}

Production cross sections of $^{147}$Pm

The radionuclide $^{147}$Pm was not quantified due to its long half-life (2.6234 a), the low abundance of the 121 keV $\gamma$-line (0.00285 \%) and the overlapping $\gamma$-lines from long lived Eu radio-products.

Production cross sections of $^{146}$Pm

The long-lived $^{146}$Pm (GS, E(level) = 0.0 keV,  $J^\pi = 3^-$,  $T_{1/2}$ = 5.53 a, $\varepsilon$: 66.0 13 \%) can only be produced by (p,2pxn) reactions directly. The excitation function is shown in Fig. 20. The agreement is best with the TENDL-2013 values. The results of the different models show large variations both in shape and in relative values. 

\begin{figure}
\includegraphics[scale=0.3]{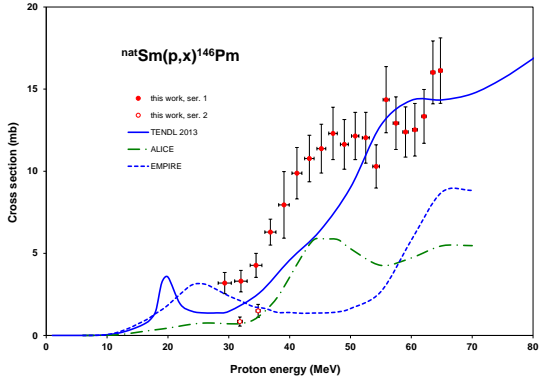}
\caption{Excitation functions of the $^{nat}$Sm(p,x)$^{146}$Pm reaction in comparison with results from model calculations}
\end{figure}

Production cross sections of $^{145}$Pm

The $\gamma$-lines of $^{145}$Pm were not detected due to the long half-life (17.7 a) and the low abundance of low intensity $\gamma$-lines and complex X-ray spectra. The results found in the TENDL-2013 library are shown in Fig 21.

\begin{figure}
\includegraphics[scale=0.3]{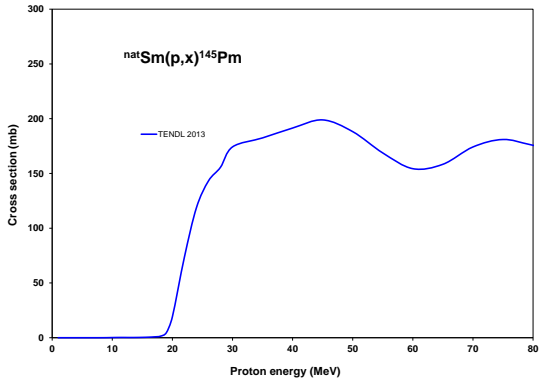}
\caption{Excitation functions of the $^{nat}$Sm(p,x)$^{145}$Pm reaction from model calculations}
\end{figure}

Production cross sections of $^{144}$Pm

The experimental and theoretical data for production of the radioisotope $^{144}$Pm (GS, E(level) = 0.0 keV,  $J^\pi = 5^-$, $T_{1/2}$ =363 d,  $\varepsilon$: 100 \%) are shown in Fig. 22.  As $^{144}$Pm is a closed nuclei there is no parent decay possible. Except for the low energy peak the agreement is best with the TENDL-2013 values. Similar questionable low energy peaks were observed in the TENDL data for different previous reactions (Figs 15-20). There are large disagreements with the data of EMPIRE at low energies and with ALICE results at high energies. The measured one point of Milazzo-Coli et al. \cite{23} is larger than our result, probably because of the unknown enrichment ratio of  their measurement ($^{147}$Sm) (see also in the case of $^{150}$Pm).

\begin{figure}
\includegraphics[scale=0.3]{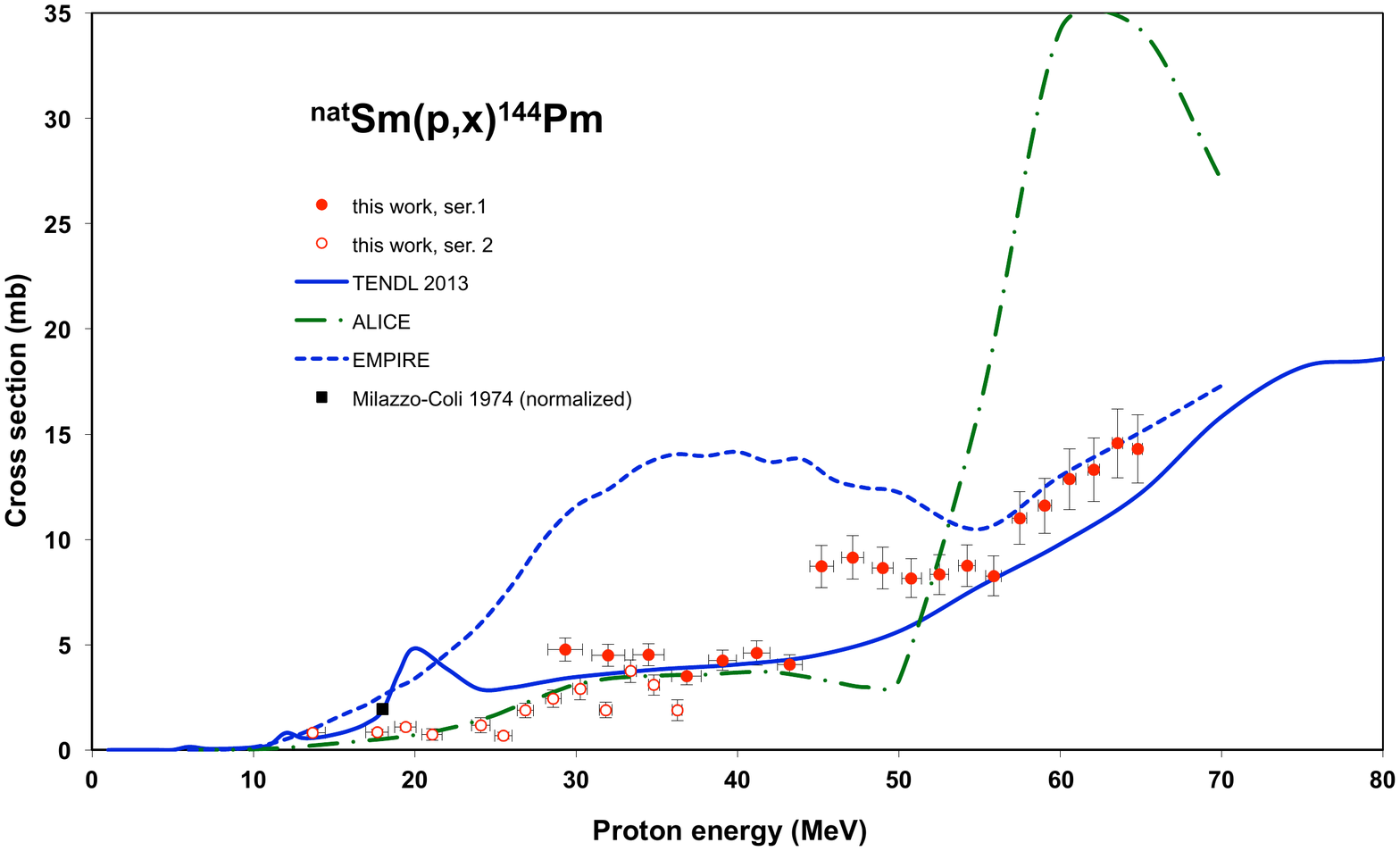}
\caption{Excitation functions of the $^{nat}$Sm(p,x)$^{144}$Pm reaction in comparison with results from model calculations}
\end{figure}

Production cross sections of $^{143}$Pm

 Our experimental cross sections for production of $^{143}$Pm (GS,  E(level) = 0.0  keV, $J^\pi = {5/2}^+$,  $T_{1/2}$ = 265 d, $\varepsilon$: 100 \%) contain the direct reaction and the total contribution from the two short-lived parent $^{143}$Sm states that we could not asses in this study, 143mSm (MS, E(level) = 753.9916 keV, $J^\pi = {11/2}^-$, $T_{1/2}$ = 66 s,  $\varepsilon$: 0.24 \%, IT: 99.76  \%) and $^{143g}$Sm (GS, E(level) = 0.0 keV, $J^\pi = {3/2}^+$, $T_{1/2}$ =  8.75 min,  $\varepsilon$: 100 \%) (Fig. 23). The comparison with the TEND3-2013 results shows a good agreement. The EMPIRE and ALICE do not present the structure of the experimental data representing the reactions on $^{144}$Sm and $^{147}$Sm.

\begin{figure}
\includegraphics[scale=0.3]{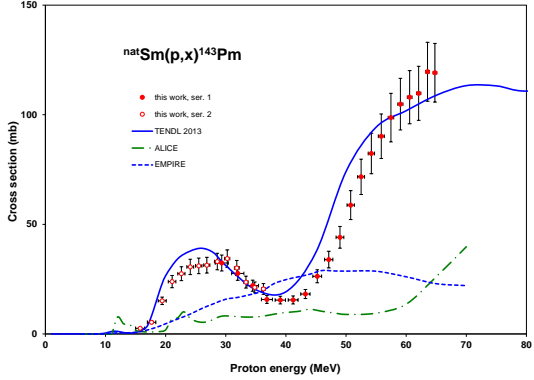}
\caption{Excitation functions of the $^{nat}$Sm(p,x)$^{143}$Pm reaction in comparison with results from model calculations}
\end{figure}

\subsubsection{Production of radioisotopes of neodymium}
\label{4.1.4}

Production cross sections of $^{141}$Nd

Our  experimental  cross section data for  $^{141}$Nd (GS,  E(level) = 0.0  keV, $J^\pi = {3/2}^+$, 2.43 h, $\varepsilon$: 100 \%) contain the direct production and the total contribution from the short-lived parent decay chain of $^{141}$Eu  (GS,  E(level) =0.0 keV, $J^\pi = {5/2}^+$, $T_{1/2}$ = 40.7 s, $\varepsilon$: 100 \%)  $^{141m}$Eu (MS,  E(level) = 96.457  keV,	  $J^\pi = {11/2}^-$, $T_{1/2}$ = 2.7 s, $\varepsilon$: 13 \%, IT: 87\%), $^{141g}$Sm  (GS,  E(level) = 0.0  keV, $J^\pi = {1/2}^+$,  $T_{1/2}$ =  10.2 min, $\varepsilon$: 100 \%), $^{141m}$Sm (E(level) = 0.0 keV, $J^\pi = {11/2}^-$,  $T_{1/2}$ = 22.6 min, IT: 0.31 \%,  $\varepsilon$: 99.69 \%) and $^{141}$Pm ((E(level) = 0.0 keV, $J^\pi = {5/2}^+$,  $T_{1/2}$ = 20.90 min, $\varepsilon$: 100 \%)  that we could not asses in this study. Only in the measurements at higher energies was the statistics sufficient to deduce cross sections, but only with high uncertainties (Fig. 24). 

\begin{figure}
\includegraphics[scale=0.3]{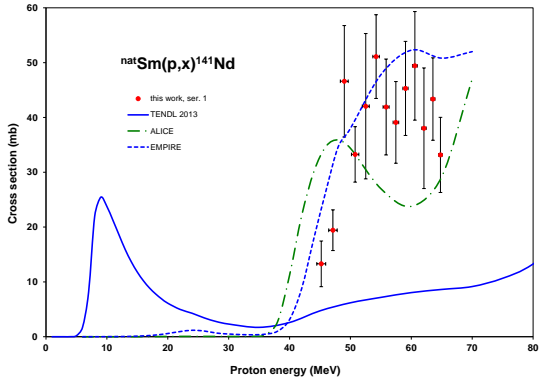}
\caption{Excitation functions of the $^{nat}$Sm(p,x)$^{141}$Nd reaction in comparison with results from model calculations}
\end{figure}

\begin{table*}[t]
\tiny
\caption{Measured cross sections of the europium radioisotopes}
\begin{center}
\begin{tabular}{|l|l|l|l|l|l|l|l|l|l|l|l|l|l|l|l|l|l|l|l|l|l|l|l|}
\hline
\multicolumn{2}{|c|}{\textbf{E$\pm\Delta$E\newline (KeV)}} & \multicolumn{2}{|c|}{\textbf{$^{154}$Eu}} & \multicolumn{2}{|c|}{\textbf{$^{152m2}$Eu}} & \multicolumn{2}{|c|}{\textbf{$^{152m1}$Eu}} & \multicolumn{2}{|c|}{\textbf{$^{
152g}$Eu}} & \multicolumn{2}{|c|}{\textbf{$^{150m}$Eu}} & \multicolumn{2}{|c|}{\textbf{$^{150g}$Eu}} & \multicolumn{2}{|c|}{\textbf{$^{149}$Eu}} & \multicolumn{2}{|c|}{\textbf{$^{148}$Eu}} & \multicolumn{2}{|c|}{\textbf{$^{147}$Eu}} & \multicolumn{2}{|c|}{\textbf{$^{146}$Eu}} & \multicolumn{2}{|c|}{\textbf{$^{145}$Eu}} \\
\hline
 \multicolumn{2}{|c|}{ } & \multicolumn{22}{|c|}{\textbf{$\sigma\pm\Delta\sigma$ mbarn}} \\
\hline
6.0 & 1.0 &  & &  & & 2.9 & 0.3 &  & & 1.0 & 0.1 &  & & 4.4 & 0.5 & 
2.1 & 0.2 & 2.7 & 0.3 &  & &  & \\
\hline
8.4 & 0.9 &  & & 0.1 & 0.01 & 16.5 & 1.9 &  & & 6.4 & 0.7 &  & & 23.2 
& 2.6 & 14.5 & 1.6 & 19.8 & 2.2 &  & &  & \\
\hline
11.3 & 0.8 &  & & 0.2 & 0.03 & 15.7 & 1.8 &  & & 8.3 & 1.0 & 17.0 & 
3.0 & 54.7 & 6.2 & 63.2 & 7.1 & 66.7 & 7.5 & 7.6 & 0.9 &  & \\
\hline
13.7 & 0.8 &  & & 0.2 & 0.03 & 6.3 & 0.8 &  & & 2.1 & 0.4 &  & & 63.0 
& 7.1 & 93.9 & 10.5 & 92.4 & 10.4 & 75.2 & 8.5 &  & \\
\hline
15.8 & 0.8 &  & & 0.2 & 0.02 & 3.6 & 0.6 &  & & 0.9 & 0.3 &  & & 69.2 
& 7.8 & 108.3 & 12.2 & 94.4 & 10.6 & 113.2 & 12.7 &  & \\
\hline
17.7 & 0.7 &  & & 0.3 & 0.04 & 4.7 & 0.7 &  & & 1.9 & 0.4 &  & & 78.5 
& 8.9 & 119.4 & 13.4 & 104.9 & 11.8 & 140.3 & 15.8 &  & \\
\hline
19.4 & 0.7 & 12.1 & 4.0 & 1.2 & 0.1 & 16.3 & 1.9 & 138.3 & 18.2 & 19.6 & 
2.3 & 18.1 & 3.6 & 77.0 & 8.7 & 111.7 & 12.5 & 121.1 & 13.6 & 147.8 & 
16.6 & 3.2 & 0.4 \\
\hline
21.1 & 0.6 &  & & 2.6 & 0.3 & 30.2 & 3.4 &  & & 42.7 & 4.8 & 73.4 & 
14.4 & 66.7 & 8.0 & 102.6 & 11.5 & 180.9 & 20.3 & 144.3 & 16.2 & 27.6 & 
3.1 \\
\hline
22.6 & 0.6 & 12.8 & 2.0 & 3.6 & 0.4 & 34.1 & 3.9 & 156.2 & 17.8 & 60.3 & 
6.8 & 109.0 & 15.5 & 39.6 & 4.9 & 93.6 & 10.5 & 173.8 & 19.5 & 142.7 & 
16.0 & 58.8 & 6.6 \\
\hline
24.1 & 0.6 & 14.9 & 2.2 & 4.6 & 0.5 & 36.8 & 4.2 & 174.7 & 19.9 & 66.5 & 
7.5 & 128.2 & 17.2 & 28.6 & 3.7 & 92.2 & 10.4 & 165.0 & 18.6 & 141.6 & 
15.9 & 82.8 & 9.3 \\
\hline
25.5 & 0.5 & 10.2 & 1.8 & 5.5 & 0.6 & 34.8 & 3.9 & 192.6 & 22.0 & 69.1 & 
7.8 & 181.4 & 23.3 & 27.2 & 3.7 & 89.2 & 10.0 & 158.6 & 17.8 & 141.6 & 
15.9 & 102.9 & 11.6 \\
\hline
26.9 & 0.5 & 12.1 & 2.1 & 5.5 & 0.6 & 29.8 & 3.4 & 172.9 & 20.0 & 61.3 & 
6.9 & 167.8 & 21.1 & 40.7 & 5.0 & 84.5 & 9.5 & 154.7 & 17.4 & 135.0 & 
15.2 & 109.8 & 12.3 \\
\hline
28.6 & 0.5 & 13.7 & 2.1 & 5.3 & 0.6 & 22.0 & 2.5 & 159.1 & 18.3 & 49.0 & 
5.5 & 150.8 & 19.1 & 81.5 & 9.4 & 80.2 & 9.0 & 150.7 & 17.0 & 146.3 & 
16.4 & 120.0 & 13.5 \\
\hline
29.3 & 1.1 &  & & 3.4 & 0.4 & 17.0 & 2.1 & 144.1 & 16.5 & 31.5 & 3.7 & 
138.5 & 15.7 & 142.5 & 16.6 & 72.8 & 8.2 & 159.7 & 18.0 & 142.4 & 16.0 & 
117.8 & 13.2 \\
\hline
30.2 & 0.4 & 13.0 & 2.0 & 4.7 & 0.5 & 17.5 & 2.0 & 117.1 & 13.5 & 38.1 & 
4.3 & 126.0 & 17.6 & 138.6 & 15.8 & 73.2 & 8.2 & 152.1 & 17.1 & 172.7 & 
19.4 & 142.3 & 16.0 \\
\hline
31.8 & 0.4 & 10.2 & 1.4 & 3.1 & 0.4 & 10.9 & 1.3 & 105.6 & 12.0 & 26.1 & 
3.0 & 99.7 & 13.9 & 177.9 & 20.1 & 56.0 & 6.3 & 132.0 & 14.9 & 172.0 & 
19.3 & 145.6 & 16.4 \\
\hline
32.0 & 1.0 &  & & 2.5 & 0.3 & 10.3 & 1.4 & 94.7 & 11.2 & 24.0 & 2.9 & 
88.8 & 10.3 & 195.4 & 23.1 & 56.5 & 6.4 & 133.9 & 15.1 & 155.4 & 17.5 & 
134.7 & 15.1 \\
\hline
33.4 & 0.4 & 7.3 & 1.4 & 2.6 & 0.3 & 8.9 & 1.1 & 52.0 & 6.2 & 20.4 & 2.4 
& 65.2 & 10.5 & 202.7 & 22.9 & 43.9 & 4.9 & 116.2 & 13.1 & 166.7 & 18.7 
& 148.4 & 16.7 \\
\hline
34.5 & 1.0 &  & & 1.8 & 0.2 & 10.8 & 1.5 & 59.6 & 8.5 & 15.9 & 2.0 & 
60.9 & 7.9 & 225.3 & 26.3 & 45.7 & 5.2 & 120.3 & 13.6 & 156.7 & 17.6 & 
139.0 & 15.6 \\
\hline
34.8 & 0.3 & 10.7 & 1.7 & 2.0 & 0.2 & 7.4 & 1.0 & 62.9 & 7.5 & 16.5 & 
2.0 & 49.7 & 11.8 & 277.1 & 31.3 & 38.1 & 4.3 & 110.7 & 12.5 & 166.9 & 
18.7 & 150.0 & 16.9 \\
\hline
36.3 & 0.3 & 9.2 & 1.6 & 1.8 & 0.2 & 7.5 & 1.0 & 55.3 & 6.3 & 20.2 & 2.4 
& 30.7 & 11.0 & 246.3 & 27.9 & 42.1 & 4.7 & 110.9 & 12.5 & 173.7 & 19.5 
& 152.4 & 17.1 \\
\hline
36.8 & 0.9 &  & & 1.2 & 0.1 & 8.0 & 1.1 & 58.1 & 7.5 & 18.1 & 2.2 & 
53.0 & 6.4 & 218.8 & 25.5 & 55.6 & 6.3 & 110.8 & 12.5 & 150.9 & 16.9 & 
130.4 & 14.7 \\
\hline
39.1 & 0.9 &  & & 1.2 & 0.1 & 7.8 & 1.1 & 65.5 & 8.1 & 23.3 & 2.8 & 
63.8 & 7.6 & 224.3 & 25.9 & 83.1 & 9.3 & 98.5 & 11.2 & 142.4 & 16.0 & 
129.8 & 14.6 \\
\hline
41.2 & 0.8 &  & & 1.2 & 0.2 & 7.4 & 1.1 & 62.3 & 14.3 & 25.7 & 3.1 & 
98.6 & 13.1 & 183.0 & 21.5 & 123.4 & 13.9 & 96.3 & 10.9 & 142.0 & 15.9 & 
140.6 & 15.8 \\
\hline
43.2 & 0.8 &  & & 1.1 & 0.1 & 5.4 & 0.9 & 22.7 & 22.8 & 28.4 & 3.3 & 
128.8 & 14.8 & 144.6 & 17.4 & 152.3 & 17.1 & 82.9 & 9.4 & 129.0 & 14.5 & 
140.4 & 15.8 \\
\hline
45.2 & 0.7 &  & & 1.0 & 0.1 & 4.1 & 0.8 & 27.1 & 5.2 & 30.1 & 3.5 & 
148.5 & 17.1 & 121.5 & 14.9 & 169.8 & 19.1 & 69.3 & 7.9 & 122.4 & 13.7 & 
141.9 & 15.9 \\
\hline
47.1 & 0.7 &  & & 0.9 & 0.1 & 7.9 & 1.0 & 40.4 & 6.7 & 28.1 & 3.3 & 
142.7 & 16.4 & 132.0 & 15.7 & 173.6 & 19.5 & 75.2 & 8.5 & 113.0 & 12.7 & 
139.2 & 15.6 \\
\hline
49.0 & 0.6 &  & & 1.2 & 0.2 & 4.8 & 0.9 & 33.7 & 6.0 & 22.5 & 2.7 & 
132.0 & 15.2 & 123.4 & 14.6 & 157.8 & 17.7 & 85.8 & 9.7 & 103.8 & 11.7 & 
131.2 & 14.7 \\
\hline
50.8 & 0.6 & 10.2 & 1.7 & 0.8 & 0.1 & 5.4 & 0.8 & 29.6 & 4.0 & 23.8 & 
2.8 & 125.4 & 14.3 & 151.8 & 23.8 & 154.7 & 17.4 & 98.7 & 11.7 & 97.5 & 
11.0 & 132.8 & 14.9 \\
\hline
52.5 & 0.6 &  & & 1.0 & 0.2 & 5.2 & 0.9 & 31.7 & 6.1 & 20.2 & 2.5 & 
101.2 & 11.8 & 152.7 & 18.3 & 134.8 & 15.1 & 121.0 & 13.7 & 90.3 & 10.1 
& 127.7 & 14.3 \\
\hline
54.2 & 0.5 &  & & 0.6 & 0.1 & 5.5 & 0.9 & 29.6 & 5.1 & 18.4 & 2.2 & 
90.7 & 10.5 & 163.1 & 18.8 & 123.3 & 13.8 & 138.6 & 15.6 & 83.3 & 9.4 & 
122.9 & 13.8 \\
\hline
55.9 & 0.5 &  & & 0.6 & 0.1 & 5.2 & 0.9 & 27.4 & 7.8 & 19.1 & 2.3 & 
89.5 & 11.2 & 151.4 & 17.6 & 113.3 & 12.7 & 145.0 & 16.3 & 78.3 & 8.8 & 
118.0 & 13.3 \\
\hline
57.5 & 0.5 & 6.0 & 1.5 & 0.7 & 0.1 & 3.0 & 0.7 & 29.5 & 4.9 & 18.3 & 2.2 
& 79.6 & 9.5 & 157.1 & 18.4 & 93.4 & 10.5 & 153.4 & 17.3 & 76.5 & 8.6 & 
111.8 & 12.6 \\
\hline
59.0 & 0.4 &  & & 0.5 & 0.1 & 6.1 & 0.9 & 31.9 & 4.9 & 15.1 & 1.9 & 
74.8 & 8.7 & 161.8 & 18.7 & 108.2 & 12.2 & 151.5 & 17.1 & 78.2 & 8.8 & 
105.1 & 11.8 \\
\hline
60.6 & 0.4 &  & & 0.7 & 0.1 & 4.7 & 0.8 & 33.8 & 5.7 & 14.3 & 1.8 & 
70.1 & 8.4 & 155.7 & 18.4 & 115.1 & 12.9 & 142.7 & 16.1 & 82.2 & 9.2 & 
100.3 & 11.3 \\
\hline
62.1 & 0.4 &  & &  & &  & & 20.7 & 4.2 & 12.2 & 1.6 & 65.2 & 7.8 & 
147.7 & 17.7 & 103.1 & 11.6 & 125.3 & 14.2 & 83.2 & 9.3 & 90.7 & 10.2 \\
\hline
63.6 & 0.3 & 7.5 & 1.5 & 0.5 & 0.1 &  & & 29.7 & 5.7 & 14.2 & 1.8 & 
56.5 & 6.8 & 120.3 & 14.3 & 45.5 & 5.1 & 130.2 & 14.7 & 93.6 & 10.5 & 
90.3 & 10.2 \\
\hline
64.8 & 0.3 &  & & 0.5 & 0.1 & 2.8 & 0.7 & 27.6 & 5.3 & 14.9 & 1.9 & 
54.8 & 6.9 & 124.3 & 15.1 & 50.0 & 5.6 & 113.8 & 12.9 & 97.1 & 10.9 & 
84.6 & 9.5 \\
\hline
\end{tabular}
\end{center}
\end{table*}

\begin{table*}[t]
\tiny
\caption{Measured cross sections of the samarium, promethium and neodymium radioisotopes}
\centering
\begin{center}
\begin{tabular}{|l|l|l|l|l|l|l|l|l|l|l|l|l|l|l|l|l|l|l|l|l|l|l|l|}
\hline
\multicolumn{2}{|c|}{\textbf{E$\pm\Delta$E\newline (KeV)}} & \multicolumn{2}{|c|}{\textbf{$^{153}$Sm}} & \multicolumn{2}{|c|}{\textbf{$^{145}$Sm}} & \multicolumn{2}{|c|}{\textbf{$^{151}$Pm}} & \multicolumn{2}{|c|}{\textbf{$^{150}$
Pm}} & \multicolumn{2}{|c|}{\textbf{$^{149}$Pm}} & \multicolumn{2}{|c|}{\textbf{$^{148m}$Pm}} & \multicolumn{2}{|c|}{\textbf{$^{148m}$Pm}} & \multicolumn{2}{|c|}{\textbf{$^{146}$Pm}} & \multicolumn{2}{|c|}{\textbf{$^{144}$Pm}} & \multicolumn{2}{|c|}{\textbf{$^{143}$Pm}} & \multicolumn{2}{|c|}{\textbf{$^{141}$Nd}} \\
\hline
 \multicolumn{2}{|c|}{ } & \multicolumn{22}{|c|}{\textbf{$\sigma\pm\Delta\sigma$ mbarn}} \\
\hline

6.0 & 1.0 &  &  &  & &  & &  & &  & &  & &  & &  & &  & &  & 
&  & \\
\hline
8.4 & 0.9 &  &  &  & &  & &  & &  & &  & &  & &  & &  & &  & 
&  & \\
\hline
11.3 & 0.8 & 0.2 & 0.05 &  & &  & &  & &  & &  & &  & &  & &  & 
&  & &  & \\
\hline
13.7 & 0.8 & 1.4 & 0.2 &  & & 0.2 & 0.1 &  & &  & &  & &  & &  & & 
0.8 & 0.2 &  & &  & \\
\hline
15.8 & 0.8 & 4.4 & 0.5 &  & &  & &  & &  & &  & &  & &  & & 0.0 & 
0.0 & 2.6 & 0.3 &  & \\
\hline
17.7 & 0.7 & 9.2 & 1.0 &  & & 0.5 & 0.1 &  & &  & &  & &  & &  & & 
0.8 & 0.1 & 5.3 & 0.6 &  & \\
\hline
19.4 & 0.7 & 13.7 & 1.5 & 1.9 & 0.5 & 0.5 & 0.1 & 0.5 & 0.1 &  & &  & 
&  & &  & & 1.1 & 0.2 & 15.1 & 1.7 &  & \\
\hline
21.1 & 0.6 & 18.8 & 2.1 & 27.2 & 4.1 & 0.7 & 0.1 & 0.7 & 0.1 &  & &  & 
&  & &  & & 0.7 & 0.3 & 23.9 & 2.8 &  & \\
\hline
22.6 & 0.6 & 21.6 & 2.4 & 57.0 & 6.8 & 0.8 & 0.1 & 1.2 & 0.1 &  & &  & 
&  & &  & & 0.0 & 0.0 & 27.5 & 3.2 &  & \\
\hline
24.1 & 0.6 & 24.0 & 2.7 & 74.7 & 8.7 & 0.6 & 0.1 & 1.5 & 0.2 &  & &  & 
&  & &  & & 1.2 & 0.4 & 30.6 & 3.5 &  & \\
\hline
25.5 & 0.5 & 26.5 & 3.0 & 96.2 & 11.1 & 1.0 & 0.2 & 1.9 & 0.2 &  & &  
& &  & &  & & 0.7 & 0.2 & 31.1 & 3.6 &  & \\
\hline
26.9 & 0.5 & 26.6 & 3.0 & 111.8 & 12.7 & 0.5 & 0.1 & 1.3 & 0.2 &  & &  
& &  & &  & & 1.9 & 0.3 & 31.4 & 3.6 &  & \\
\hline
28.6 & 0.5 & 28.4 & 3.2 & 127.2 & 14.5 & 0.7 & 0.1 & 1.4 & 0.2 &  & &  
& &  & &  & & 2.4 & 0.4 & 33.0 & 3.8 &  & \\
\hline
29.3 & 1.1 & 25.8 & 2.9 & 111.7 & 12.5 &  & & 1.7 & 0.4 &  & & 0.5 & 
1.0 &  & & 3.2 & 0.6 & 4.8 & 0.5 & 32.4 & 3.6 &  & \\
\hline
30.2 & 0.4 & 31.1 & 3.5 & 141.0 & 16.1 &  & & 2.0 & 0.3 &  & &  & &  
& &  & & 2.9 & 0.5 & 34.4 & 4.0 &  & \\
\hline
31.8 & 0.4 & 29.9 & 3.4 & 156.5 & 17.7 &  & & 1.6 & 0.2 &  & &  & &  
& & 0.8 & 0.3 & 1.9 & 0.4 & 30.1 & 3.5 &  & \\
\hline
32.0 & 1.0 & 26.4 & 3.0 & 123.3 & 13.9 &  & & 1.8 & 0.5 &  & &  & &  
& & 3.3 & 0.7 & 4.5 & 0.5 & 27.6 & 3.1 &  & \\
\hline
33.4 & 0.4 & 28.8 & 3.2 & 150.6 & 17.1 &  & & 1.6 & 0.2 &  & &  & &  
& &  & & 3.7 & 0.5 & 23.6 & 2.7 &  & \\
\hline
34.5 & 1.0 & 26.5 & 3.0 & 144.0 & 16.2 &  & & 1.6 & 1.0 &  & &  & &  
& & 4.3 & 0.7 & 4.5 & 0.5 & 22.0 & 2.5 &  & \\
\hline
34.8 & 0.3 & 29.9 & 3.4 & 162.3 & 18.4 &  & & 1.7 & 0.3 &  & &  & &  
& & 1.5 & 0.4 & 3.1 & 0.5 & 21.2 & 2.5 &  & \\
\hline
36.3 & 0.3 & 32.2 & 3.6 & 162.7 & 18.5 &  & & 1.2 & 0.2 &  & &  & &  
& &  & & 1.9 & 0.5 & 20.5 & 2.5 &  & \\
\hline
36.8 & 0.9 & 27.5 & 3.1 & 127.9 & 14.4 & 0.8 & 0.2 & 1.0 & 0.2 &  & &  
& &  & & 6.3 & 0.8 & 3.5 & 0.4 & 15.7 & 1.8 &  & \\
\hline
39.1 & 0.9 & 27.2 & 3.1 & 169.3 & 19.0 &  & & 1.1 & 0.2 &  & &  & &  
& & 8.0 & 2.0 & 4.3 & 0.5 & 15.4 & 1.7 &  & \\
\hline
41.2 & 0.8 & 29.5 & 3.3 & 117.4 & 13.2 & 0.8 & 0.2 & 1.4 & 0.3 &  & & 
3.1 & 0.9 &  & & 9.9 & 1.6 & 4.6 & 0.6 & 15.5 & 1.8 &  & \\
\hline
43.2 & 0.8 & 29.1 & 3.3 & 160.1 & 18.0 & 0.5 & 0.1 & 1.3 & 0.3 &  & & 
5.1 & 1.1 & 1.1 & 0.2 & 10.8 & 1.4 & 4.1 & 0.5 & 18.2 & 2.1 &  & \\
\hline
45.2 & 0.7 & 28.8 & 3.2 & 139.4 & 15.7 & 0.4 & 0.1 & 1.1 & 0.2 &  & & 
4.9 & 1.1 &  & & 11.4 & 1.5 & 8.7 & 1.0 & 26.3 & 3.0 & 13.3 & 4.2 \\
\hline
47.1 & 0.7 & 28.9 & 3.3 & 149.0 & 16.7 & 0.7 & 0.2 & 1.2 & 0.2 &  & & 
3.1 & 0.7 &  & & 12.3 & 1.6 & 9.1 & 1.0 & 33.9 & 3.8 & 19.4 & 3.7 \\
\hline
49.0 & 0.6 & 28.2 & 3.2 & 164.3 & 18.5 & 0.8 & 0.2 & 1.5 & 0.4 &  & & 
4.0 & 0.9 &  & & 11.6 & 1.5 & 8.6 & 1.0 & 44.1 & 5.0 & 46.6 & 10.2 \\
\hline
50.8 & 0.6 & 29.0 & 3.3 & 161.1 & 18.1 & 1.1 & 0.2 & 1.9 & 0.3 & 2.8 & 
0.7 &  & &  & & 12.1 & 1.4 & 8.1 & 0.9 & 58.8 & 6.6 & 33.3 & 5.1 \\
\hline
52.5 & 0.6 & 28.8 & 3.2 & 157.6 & 17.7 & 0.9 & 0.2 & 1.5 & 0.3 &  & & 
4.3 & 1.0 & 0.7 & 0.2 & 12.0 & 1.5 & 8.3 & 0.9 & 71.7 & 8.1 & 42.0 & 
13.3 \\
\hline
54.2 & 0.5 & 29.2 & 3.3 & 144.9 & 16.3 & 1.5 & 0.2 & 1.3 & 0.2 &  & & 
5.4 & 0.8 &  & & 10.3 & 1.3 & 8.7 & 1.0 & 82.3 & 9.3 & 51.1 & 7.6 \\
\hline
55.9 & 0.5 & 29.3 & 3.3 & 155.3 & 17.5 & 1.4 & 0.2 & 1.7 & 0.3 &  & & 
4.6 & 0.9 & 0.2 & 0.2 & 14.4 & 2.0 & 8.3 & 0.9 & 90.2 & 10.2 & 41.9 & 
8.8 \\
\hline
57.5 & 0.5 & 28.9 & 3.3 & 164.1 & 18.4 & 1.2 & 0.2 & 2.5 & 0.4 & 2.6 & 
0.8 & 2.7 & 0.7 & 0.5 & 0.2 & 12.9 & 1.6 & 11.0 & 1.2 & 98.7 & 11.1 & 
39.1 & 7.5 \\
\hline
59.0 & 0.4 & 28.7 & 3.2 & 153.4 & 17.2 & 1.9 & 0.3 & 2.1 & 0.4 &  & & 
4.2 & 0.8 & 0.7 & 0.1 & 12.4 & 1.5 & 11.6 & 1.3 & 104.8 & 11.8 & 45.3 & 
8.6 \\
\hline
60.6 & 0.4 & 28.6 & 3.2 & 156.5 & 17.6 & 1.7 & 0.2 & 2.2 & 0.4 &  & &  
& & 0.4 & 0.3 & 12.5 & 1.6 & 12.9 & 1.5 & 108.0 & 12.1 & 49.4 & 9.9 \\
\hline
62.1 & 0.4 & 28.0 & 3.2 & 133.7 & 15.0 & 1.1 & 0.2 & 1.7 & 0.5 &  & & 
1.2 & 0.5 & 0.6 & 0.1 & 13.3 & 1.6 & 13.3 & 1.5 & 109.8 & 12.3 & 38.0 & 
11.0 \\
\hline
63.6 & 0.3 & 28.5 & 3.2 & 121.6 & 13.7 & 1.6 & 0.2 & 2.0 & 0.3 &  & &  
& & 0.7 & 0.1 & 16.0 & 1.9 & 14.6 & 1.6 & 119.6 & 13.4 & 43.4 & 7.5 \\
\hline
64.8 & 0.3 & 27.9 & 3.1 & 119.1 & 13.4 & 2.0 & 0.3 & 1.8 & 0.3 &  & & 
0.9 & 0.4 & 0.7 & 0.2 & 16.1 & 2.0 & 14.3 & 1.6 & 119.2 & 13.4 & 33.2 & 
6.9 \\
\hline
\end{tabular}
\end{center}
\end{table*}

\section{Thick target yields}
\label{5}

Thick target yields (integrated yield for a given incident energy down to the reaction threshold) were calculated from fitted curves to our experimental cross section data for the investigated radionuclides. The results for physical yields \cite{17} (no saturation effects, obtained in an instantaneous short irradiation time for particles having a charge equivalent to 1 µAh) are presented in figure 25-26 and compared to experimentally measured yields of Dmitriev et al. \cite{7}. Our calculated data are in very good agreement with the data of Dmitriev in the case of $^{147,152g}$Eu, and give slightly higher values in the case of $^{148}$Eu. There are no literature value found for the cases of Sm, Pm and Nd isotopes. 

\begin{figure}
\includegraphics[scale=0.3]{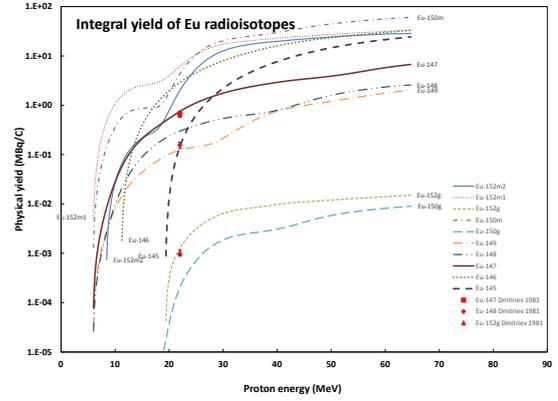}
\caption{Integral thick target yields for the formation of $^{152m2,152m1,152g,150m,150g,149,148,147,146,145}$Eu in proton induced nuclear reaction on $^{nat}$Sm as function of the proton energy.}
\end{figure}

\begin{figure}
\includegraphics[scale=0.3]{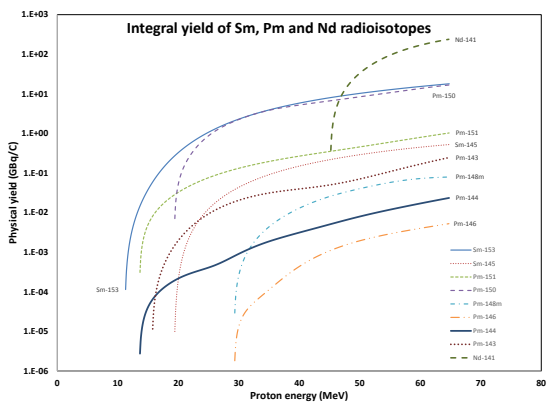}
\caption{Integral thick target yields for the formation of $^{153,145}$Sm, $^{151,150,148m,146,144,143}$Pm and $^{141}$Nd  as function of the proton energy}
\end{figure}

\section{Comparison of production routes of $^{145}$Sm and $^{153}$Sm.}
\label{6}

The comparison of the production routes of these two radiolanthanides was discussed in detail in our previous work on deuteron induced activation data (for details see \cite{3}). Among the competitive production routes we discussed only the low energy, light charged particle induced routes including proton-induced reactions on some samarium isotopes. In the reference the activation data were based on theoretical results (based on the nuclear model codes, see Table 1) due to the lack of experimental data at that time. The present experimental data for production of $^{145}$Sm (and its parent $^{145}$Eu) and $^{153}$Sm support the results of the theoretical estimations for $^{nat}$Sm target. We can expect with high probability that also the reaction cross sections for the contribution on selected stable Sm target isotopes are confirmed and that the conclusions of the earlier paper remains unchanged. 
“The route to get carrier free, high specific activity $^{145}$Sm is possible only with accelerators through the decay of $^{145}$Eu or through the direct production from alpha induced nuclear reactions on natural or enriched neodymium.  Indirect production through $^{145}$Eu obtained from irradiation of Sm targets asks for two chemical separations while in case of direct production one chemical separation is needed to produce a high-specific-activity and high radionuclidic purity $^{145}$Sm end product.
Out of them the $^{147}$Sm(p,3n)$^{145}$Eu-$^{145}$Sm is the most productive ($^{147}$Sm 15 \% in natural) and asks for 30 MeV machines.
For production of $^{153}$Sm it is difficult to compete for efficiency with the reactors due to the large neutron cross section and because the products of the proton induced reactions are also carrier added. The proton induced reactions can however have importance for local use due to the relatively short half-life of $^{153}$Sm ($T_{1/2}$ = 46.50 h) and due to the free capacity of charged particle accelerators.	

\section{Comparison of production routes of $^{145}$Sm and $^{153}$Sm.}
\label{7}
In the frame of a systematic investigation of low and medium energy excitation functions of activation products for different applications, the proton induced cross sections on samarium were investigated for the first time. Independent or cumulative cross-sections for the formation of the radionuclides $^{154,152m2,152m1,152g,150m,150g,149,148,147,146,145}$Eu, $^{153,145}$Sm, $^{151,150,149,148g,148m,146,144,143}$Pm and $^{141}$Nd up to 65 MeV through $^{nat}$Sm(p,x) nuclear reactions were measured up to 50 MeV for the first time.
The experimental data were compared with the theoretical predictions obtained by our EMPIRE and ALICE-IPPE code calculations and with the TALYS data in the TENDL-2013 library. The predictions of the different codes differ significantly from each other and from the experimental data. As samarium is an important technological material, the measured activation cross sections can be useful in the field of medical isotope production and to estimate the radiation dose caused by primary or secondary proton particles in reactors, accelerators and in space.

\section{Acknowledgements}

This work was performed in the frame of the HAS-FWO Vlaanderen (Hungary-Belgium) project. The authors acknowledge the support of the research project and of the respective institutions in providing the beam time and experimental facilities

 



\bibliographystyle{elsarticle-num}
\bibliography{Smp}

\begin{thebibliography}{10}
\expandafter\ifx\csname url\endcsname\relax
  \def\url#1{\texttt{#1}}\fi
\expandafter\ifx\csname urlprefix\endcsname\relax\def\urlprefix{URL }\fi
\expandafter\ifx\csname href\endcsname\relax
  \def\href#1#2{#2} \def\path#1{#1}\fi

\bibitem{1}
IAEA, Manual for Reactor Produced Radioisotopes, IAEA TECDOC 1340, Vol. 1340,
  IAEA, Vienna, 2003.

\bibitem{2}
IAEA, Nuclear data for the production of therapeutic radionuclides. —
  Technical reports series, no. 473, IAEA, Vienna, 2011.

\bibitem{3}
F.~T\'ark\'anyi, A.~Hermanne, S.~Tak\'acs, F.~Ditr\'oi, J.~Csikai, A.~V.
  Ignatyuk, Cross sections of deuteron induced reactions on sm-nat for
  production of the therapeutic radionuclide sm-145 and sm-153, Applied
  Radiation and Isotopes 91 (2014) 31--37.

\bibitem{4}
F.~T\'ark\'anyi, A.~Hermanne, S.~Tak\'acs, F.~Ditr\'oi, J.~Csikai, A.~V.
  Ignatyuk, Activation cross-sections of deuteron induced reactions on sm-nat
  up to 50 mev, Nuclear Instruments \& Methods in Physics Research Section
  B-Beam Interactions with Materials and Atoms 333 (2014) 12--26.

\bibitem{5}
D.~Updegraff, S.~A. Hoedl, Nuclear medicine without nuclear reactors (2013).

\bibitem{6}
Laka-Foundation, Medical radioisotope production without a nuclear reactor
  (2010).

\bibitem{7}
P.~P. Dmitriev, G.~A. Molin, Radioactive nuclide yields for thick target at 22
  mev proton energy, Vop. At. Nauki i Tekhn., Ser.Yadernye Konstanty 44~(5)
  (1981) 43.

\bibitem{8}
NuDat, Nudat2 database (2.6) (2014).

\bibitem{9}
B.~Pritychenko, A.~Sonzogni, Q-value calculator (2003).

\bibitem{10}
{International-Bureau-of-Weights-and-Measures}, Guide to the expression of
  uncertainty in measurement, 1st Edition, International Organization for
  Standardization, Genève, Switzerland, 1993.

\bibitem{11}
Canberra,
  http://www.canberra.com/products/radiochemistry\_lab/genie-2000-software.asp.
  (2000).

\bibitem{12}
G.~Sz\'ekely, Fgm - a flexible gamma-spectrum analysis program for a small
  computer, Computer Physics Communications 34~(3) (1985) 313--324.

\bibitem{13}
F.~T\'ark\'anyi, F.~Szelecs\'enyi, S.~Tak\'acs, Determination of effective
  bombarding energies and fluxes using improved stacked-foil technique, Acta
  Radiologica, Supplementum 376 (1991) 72.

\bibitem{14}
H.~H. Andersen, J.~F. Ziegler, Hydrogen stopping powers and ranges in all
  elements. The stopping and ranges of ions in matter, Volume 3., The Stopping
  and ranges of ions in matter, Pergamon Press, New York, 1977.

\bibitem{15}
T.~Belgya, O.~Bersillon, R.~Capote, T.~Fukahori, G.~Zhigang, S.~Goriely,
  M.~Herman, A.~V. Ignatyuk, S.~Kailas, A.~Koning, P.~Oblozinsky, V.~Plujko,
  P.~Young, Handbook for calculations of nuclear reaction data: Reference Input
  Parameter Library. http://www-nds.iaea.org/RIPL-2/, IAEA, Vienna, 2005.

\bibitem{16}
F.~T\'ark\'anyi, S.~Tak\'acs, K.~Gul, A.~Hermanne, M.~G. Mustafa, M.~Nortier,
  P.~Oblozinsky, S.~M. Qaim, B.~Scholten, Y.~N. Shubin, Z.~Youxiang, Tecdoc
  1211, iaea, 2001 beam monitor reactions (chapter 4). charged particle
  cross-section database for medical radioisotope production: diagnostic
  radioisotopes and monitor reactions., Tech. rep., IAEA (2001).

\bibitem{17}
M.~Bonardi, The contribution to nuclear data for biomedical radioisotope
  production from the milan cyclotron facility (1987).

\bibitem{18}
A.~I. Dityuk, A.~Y. Konobeyev, V.~P. Lunev, Y.~N. Shubin, New version of the
  advanced computer code alice-ippe, Tech. rep., IAEA (1998).

\bibitem{19}
M.~Herman, R.~Capote, B.~V. Carlson, P.~Oblozinsky, M.~Sin, A.~Trkov,
  H.~Wienke, V.~Zerkin, Empire: Nuclear reaction model code system for data
  evaluation, Nuclear Data Sheets 108~(12) (2007) 2655--2715.

\bibitem{20}
A.~J. Koning, D.~Rochman, S.~van~der Marck, J.~Kopecky, J.~C. Sublet, S.~Pomp,
  H.~Sjostrand, R.~Forrest, E.~Bauge, H.~Henriksson, O.~Cabellos, S.~Goriely,
  J.~Leppanen, H.~Leeb, A.~Plompen, R.~Mills, Tendl-2013: Talys-based evaluated
  nuclear data library (2012).

\bibitem{21}
A.~J. Koning, S.~Hilaire, M.~C. Duijvestijn, Talys-1.0 (2007).

\bibitem{22}
I.~Gheorghe, D.~Filipescu, T.~Glodariu, D.~Bucurescu, I.~Cata-Danil,
  G.~Cata-Danil, D.~Delenau, D.~Ghita, M.~Ivascu, R.~Lica, N.~Marginean,
  R.~Marginean, C.~Mihai, A.~Negret, T.~Sava, L.~Stroe, S.~Toma, O.~Sima,
  M.~Sin, Absolute cross sections for proton induced reactions on 147,149sm
  below the coulomb barrier, Nucl. Data Sheets 119 (2014) 245--248.

\bibitem{23}
L.~{Milazzo-Colli}, M.~{Braga-Marcazzan}, M.~Milazzo, G.~Signorini,
  Preformation probability of alpha-clysters in rare earth nuclei measured by
  means of the (p,a)reaction, Nuclear Physics A 218 (1974) 274.

\bibitem{24}
S.~Y.~F. Chu, L.~P. Ekström, R.~B. Firestone, Www table of radioactive
  isotopes, version 2.1 http://ie.lbl.gov/toi/ (2004).

\end{thebibliography}







\end{document}